\documentclass[preprint]{aastex}
\slugcomment{Accepted by PASP, 5 September 2012}

\shorttitle{Nova Atlas}
\shortauthors{Walter et al.}

\begin{document}
\title{The Stony Brook / SMARTS Atlas of (mostly) Southern Novae}
\author{Frederick M. Walter, Andrew Battisti\altaffilmark{1} \& Sarah E. Towers\altaffilmark{2}}
\affil{Department of Physics and Astronomy, Stony Brook University, Stony Brook, NY 11794-3800}
\author{Howard E. Bond}
\affil{Space Telescope Science Institute, Baltimore MD 21218\altaffilmark{3}}
\author{Guy S. Stringfellow}
\affil{Center for Astrophysics and Space Astronomy, University of Colorado, Boulder CO 80309}
\altaffiltext{1}{now at Dept of Astronomy, University of Massachusetts,
Amherst MA 01003}
\altaffiltext{2}{now at Dept of Physics, Western Michigan University, Kalamazoo MI, 49008}
\altaffiltext{3}{current address: 9615 Labrador Ln., Cockeysville, MD 21030}

\begin{abstract}

We introduce the Stony Brook / SMARTS Atlas of (mostly) Southern Novae.
This atlas contains both spectra and photometry obtained since 2003. 
The data archived in this atlas will facilitate systematic studies
of the nova phenomenon and correlative studies with other comprehensive
data sets. It will also enable detailed investigations of individual objects.
In making the data public we hope to engender more interest
on the part of the community in the physics of novae.
The atlas is on-line at
\url{http://www.astro.sunysb.edu/fwalter/SMARTS/NovaAtlas/}.

\end{abstract}
\section{Introduction}
Exploding stars have been noted for millennia, and observed (in a scientific
sense) for somewhat over a century. 
It wasn't until the middle of the 20$^{th}$ century that a distinction
could be made
between the supernovae, the novae, and the eruptive phenomena seen
in cataclysmic variables (the dwarf novae).
\cite{K63} was the first to suggest that the
novae were the consequence of explosive hydrogen burning on the surface
of a degenerate dwarf. It is now well accepted that
the novae are manifestations of runaway thermonuclear reactions
on the
surface of a white dwarf (WD) accreting hydrogen in a close binary system
(e.g., \citealt{S71}).
The novae are highly dynamic phenomena, with
timescales ranging from seconds to millennia, occurring in complex systems
involving two stars and mass transfer.

The primary driver of the evolution of the observational characteristics
of a nova is 
the temporal decrease of the optical depth in an expanding atmosphere.
The novae are marked by an extraordinary spectral evolution \citep{Wil91,Wil92}.
In the initial phases one often
sees an optically thick, expanding pseudo-photosphere. In some cases
one sees the growth and then disappearance of inverse P~Cygni
absorption lines from the cool,
high velocity ejecta. As the pseudo-photosphere becomes optically thin, emission
lines of the Hydrogen Balmer series strengthen, accompanied by either a
spectrum dominated by permitted lines of Fe~II, or of helium and nitrogen.
The emission line profiles and line ratios evolve as the optical depth
of the ejecta decreases, and the nova transitions from the permitted to the 
nebular phase \citep{Wil91}.

Beyond this template, in detail the novae exhibit a panoply
of individual behaviors.
\cite{PG57} and \cite{McL60}
described the evolution of novae as they were known at the time. 
\cite{Wil92} discussed the formation of the lines, and
divided novae into the \ion{Fe}{2} and He-N classes, based on 
which emission lines dominated (aside from the ubiquitous
Balmer lines of hydrogen).
Novae are also categorized as recurrent and classical novae, with the former
having more than one recorded outburst. Over a long enough baseline, it is
likely that all novae are recurrent (e.g., \citealt{For78}).

\cite{BE08} present a recent set of reviews of the nova phenomenon.

There exist well-sampled photometric records for many novae, such as those 
presented by \cite{SSH10}. They classify the photometric light curves, from 
plates amassed over the past century, into 7 distinct photometric classes.
On the other hand, spectroscopic observations of novae have rarely been 
pursued far past maximum because most novae fade rapidly, and time on the
large telescopes required for spectroscopy is precious.
The most comprehensive past work was the Tololo Nova atlas
\citep{Wil94} of 13 novae followed spectroscopically over a
5 year interval.
The availability of the SMARTS\footnote{The Small and Medium Aperture
Telescope System, directed by Charles Bailyn, is an ever-evolving
partnership that has overseen operations of 4 small telescopes at
Cerro Tololo Interamerican Observatory since 2003.}
telescope facilities \citep{S10} makes
possible routine synoptic monitoring programs, 
both photometric and spectroscopic,
of time-variable sources.
It is timely, therefore, to undertake a comprehensive, systematic,
high-cadence study of the
spectrophotometric evolution of the galactic novae.

This atlas collects photometry and spectra of the novae we have observed
with SMARTS.
Most of the novae in the atlas are recent novae, discovered since 2003. Most
are in the southern hemisphere. The observing cadences are irregular; we have
concentrated on He-N and recurrent novae, novae in the LMC, and
novae that otherwise show unusual characteristics.

Our purpose here is to introduce this atlas.
Our scientific aim is to facilitate a detailed comparison of the
various characteristics of the novae. These data can be used by and of
themselves to study individual objects, 
for systematic studies to further define the phenomenon, 
and for correlative studies with other comprehensive data sets,
such as the {\it Swift} Nova Working Group's survey of X-ray and UV
observations of recent novae \citep{Swift11}.
Our aim in making the atlas public is to make the data accessible to 
the community.
We are focusing on certain novae, 
and on particular aspects of the nova phenomenon (\S\ref{sec-disc}),
but simply cannot do justice to the full dataset.

\section{Observations and Data Analysis}

\subsection{Low Dispersion Spectroscopy}\label{sec-lds}
The spectra reported here have been obtained with the venerable RC 
spectrograph\footnote{\url{http://www.ctio.noao.edu/spectrographs/60spec/60spec.html}}
on the SMARTS/CTIO 1.5m telescope.
Observations are queue-scheduled, and are taken by
dedicated service observers.

The detector is a Loral 1K CCD. We use a variety of spectroscopic modes, with
most of the spectra having been obtained with one of the standard modes 
shown in Table~\ref{tbl-spsetups}.

We use slit widths of 1 and 0.8 arcsec in the low and the higher resolution
47/II modes, respectively.
The slit is oriented E-W and is not rotated during the night.

We routinely obtain 3 spectra of each target in order to filter for cosmic 
rays. We combine the 3 images and extract the spectrum by fitting a Gaussian 
in the spatial direction at each pixel. Wavelength calibration is accomplished 
by fitting a 3$^{rd}$ to 6$^{th}$ order polynomial to the
Th-Ar or Ne calibration lamp 
line positions. We observe a spectrophotometric standard star, generally 
LTT~4364 \citep{Ham92,Ham94} or Feige~110 \citep{Oke90,Ham92,Ham94},
on most nights to determine the counts-to-flux conversion.
Because of slit losses, possible changes in transparency and seeing 
during the night, and parallactic losses due to the fixed slit orientation,
the flux calibration is imprecise. We generally recover the correct 
spectral shape, except at the shortest wavelengths ($<$3800\AA)
where apparent changes in the slope of the continuum 
are likely attributable to airmass-dependent parallactic slit losses.
We have the capability to use simultaneous or contemporaneous photometry
to recalibrate the spectra.

There are some quality control issues that have not been fully dealt with,
especially when we are near the sensitivity
limits of the telescope. These include
observations of an incorrect star, obviously incorrect flux calibrations,
or spectra indistinguishable from noise. We are going through the data
as time permits to address these issues.

\subsection{High Dispersion Spectroscopy}

We have a small number of high resolution spectra of some of the
brighter novae near maximum.
These were obtained with the Bench-Mounted 
Echelle\footnote{\url{http://www.ctio.noao.edu/noao/content/fiber-echelle-spectrograph}}, and
currently with the Chiron echelle
spectrograph\footnote{\url{http://www.ctio.noao.edu/noao/content/chiron}}. 
These data will be incorporated into the atlas at a later time.

\subsection{Photometry}

Most of the photometry was obtained using the 
ANDICAM\footnote{\url{http://www.astronomy.ohio-state.edu/ANDICAM/detectors.html}}
dual-channel imager on the 1.3m telescope.
Observations are queue-scheduled, with dedicated service observers.

The ANDICAM optical channel is a 2048$^2$ pixel Fairchild 447 CCD.
It is read out with 2x2 binning, which yields a 0.369 arcsec/pixel
plate scale. The field of view is roughly 6x6 arcmin, but until
recently there has been 
significant unusable area on the east and south sides on the chip.
The finding charts in the atlas
show examples of ANDICAM images. We normally obtain single
images, since the fraction of pixels marred by cosmic rays
and other events is small.
Exposure times range from 1 second to about 2 minutes. We use the standard 
Johnson-Kron-Cousins $B$, $V$,
$R_C$, and $I_C$ filters (the $U$ filter has been unavailable since 2005, but
we have extensive $U$~band for some of the earlier novae, particularly
V475~Sct and V5114~Sgr).

The ANDICAM IR channel is a Rockwell 1024$^2$ HgCdTe ``Hawaii'' Array.
It is read out in 4 quadrants with 2x2 binning, which yields a
0.274 arcsec/pixel plate scale and a 2.4 arcmin field of view.
The observations are dithered using an internal mirror. In most cases
we use 3 dither positions, with integration times from 4 seconds (the minimum
integration time) to about 45 seconds. We use the CIT/CTIO $J$, $H$, and $K_s$
filter set.

The optical and IR channels are observed simultaneously using a dichroic beam
splitter.
The observing cadence varies from nightly for new novae to $\sim$annual
monitoring for the oldest novae in our list.

We perform aperture photometry on the target and between 1 and 25
comparison stars in the field. The aperture radius R is either 5 or 7
pixels, depending on field crowding and sky brightness.
The background is the median
value in an annulus of inner radius 2R and outer radius 2R+20 pixels centered
on the extraction aperture.
Instrumental magnitudes are recorded for each star. There are cases where the
fading remnant becomes blended with nearby stars (within $\sim$1.5~arcsec).
To date we have not accounted for such blending.
Eventually we plan to employ PSF-fitting techniques
in these crowded regions.

On most photometric nights an observation of a \cite{Landolt92} standard field
is taken. On those nights 
we determine the zero-point correction and determine the magnitudes of the
comparison stars. We adopt the mean magnitudes for each comparison star.
These are
generally reproducible to better than 0.02~mag; variable stars are identified
through their scatter around the mean, and are not used in the 
differential photometry.
With only a single observation of a standard star field each night,
we assume the nominal atmospheric extinction law and zero color correction. 
Using differential photometry, we can
recover the apparent magnitude of a target with 
a typical uncertainty of $<$0.03 mag at 20$^{th}$ magnitude.

While we could do the same with the IR channel images, we find it simpler to
use the catalogued 2MASS magnitudes of the standard stars. We implicitly
assume that the 2MASS comparison stars are non-variable, and that the color
terms in the photometric solution are negligible.

In addition, some higher cadence data have been obtained with the SMARTS 0.9m
and 1.0m telescopes.
The 0.9m detector is a 2048x2046 CCD with a 0.401~arcsec/pixel plate scale.
On the 1.0m, we used the 512x512 Apogee camera that was employed prior to
installation of the 4K camera.
We perform the differential photometry
in a manner identical to that for the ANDICAM, and merge the data sets.
These data are not yet fully incorporated into the atlas.

\section{Setup of the Atlas}

The atlas is on-line at
\url{http://www.astro.sunysb.edu/fwalter/SMARTS/NovaAtlas/}.

The atlas consists of a main page for each nova, giving finding charts (in
both $V$ and $K$ bands), coordinates, and links to the spectra and photometry.
The spectra are available as images, and may be downloaded in ascii (text)
format. The photometry page shows plots of the light curve and colors,
and permits one to download the data in ascii format.
Note that the plots on the photometry page only show data with formal
uncertainties $<$0.5~mag, while all measured magnitudes and uncertainties
are included in the ascii listings.
There is a link to a page of references for other observations of the novae.

\section{The Novae}

As of 1 July 2012 the atlas includes data on 64 novae. 
Of these, 29 are still bright enough (V$<$18) to reach spectroscopically
with the 1.5m/RC spectrograph. Most are still detectable photometrically
with the 1.3m/ANDICAM imager. Only 5 are no longer on our photometric
target list because
they are too faint or too confused with brighter companions.

\begin{figure}
\epsscale{0.8}
\includegraphics[scale=.8,angle=-90.]{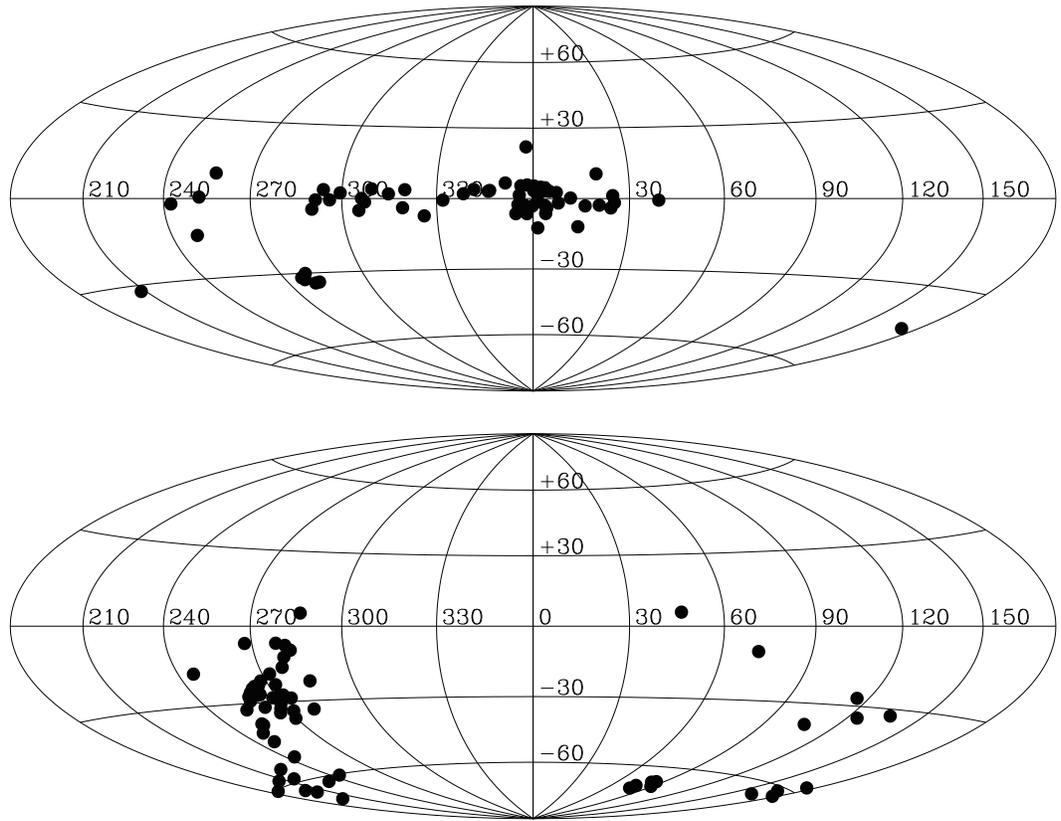}
\caption{The spatial distribution of the novae currently in the atlas.
The lower plot shows the distribution in celestial coordinates, centered at
RA=0.0; the upper plot shows the distribution in galactic coordinates, 
centered at $\ell^{II}$,B$^{II}$=0,0.
\label{fig_spdist}}
\end{figure}

The spatial distribution of these novae is shown in
Figure~\ref{fig_spdist} in both celestial and galactic
coordinates.

Lists of our targets and particulars 
on the number and observing date distribution of the 
observations are in Tables~\ref{tbl-obs} (novae from before 2012);
\ref{tbl-obs_new} (novae discovered in 2012), and \ref{tbl-obs_LMC} (novae in
the LMC). The reference time is ideally the time of peak brightness, but this
is often not well known. In general, T$_0$ is the time is discovery.
In the case of T Pyx, which rises very slowly, T$_o$ is the time of peak
brightness as estimated from our photometry. For novae that were discovered well
past peak, including N Sgr 2012b and XMMU J115113.3-623730, T$_0$ is a guess.
All the dates in the Tables are referenced to T$_0$.
The tabulated $V$ is the last observed $V$ magnitude;
in most cases this is
the brightness in June 2012.
The Tables are current as of 1 July 2012.

\normalsize
\subsection{Observing Statistics}

As of 1 July 2012 the full atlas contains 64 novae.
We have between 1 and 368 spectra for the novae, with a median of 28
spectra per nova. The number of photometric points varies between 1 and 265,
with a median of 35,
for 53 novae. Since some of the observations were taken through thick clouds,
not all observations have the best possible S/N.

The photometric and spectral coverage is generally non-uniform in time. 
In addition to annual gaps due to the Sun, there is spotty coverage during the
austral winter when the weather becomes worse. We do not have unlimited 
observing time, so we concentrate on those novae that tickle our astronomical
fancy - the He-N novae, and those showing unusual characteristics. We 
do not attempt spectroscopy of targets fainter than $V\sim$18, because
the 1.5m telescope has limited grasp. 

We generally do not make great efforts to obtain photometry from day 0, because
amateur astronomers do such a good job. In many cases data available from the
AAVSO can fill in the first few weeks, while the nova is bright (we do have
bright limits near $V=8$ and $K=6$). Our forte is the ability
to a) follow the evolution to quiescence, and b) to do so in the 7 photometric
bands from $B$ through $K_s$. 
In one case we were on the nova 1.1 days after discovery,
but the median delay is 15 days.

We try to start the spectroscopic monitoring sooner, because this is a unique
capability of SMARTS. The first spectrum is obtained with a median delay of
8.0 days from discovery, but we have observed 1 nova within 0.6~days of
discovery, 9 within 2 days, and 14 within 3 days. 

We have multi-epoch photometry of 52 novae over timespans of up to
3173 days (8.7 years),
and multi-epoch spectroscopy of 63 novae over timespans of up to
3156 days (8.6 years). These durations will increase with time so long as
SMARTS continues operating, and the targets are sufficiently bright. The median
observation durations of 1317 and 360 days, respectively, for the photometry
and spectroscopy, are limited mostly by target brightness.
The median time between observations is skewed by the growing number of old,
faint targets that are now observed with a cadence of 1-2 observations per
year, so that the median time between spectra is 8.4 days, and is 27 days
for photometric observations.

\subsection{The Example of V574 Pup}

We illustrate possible uses of the atlas with the example of V574 Pup,
an \ion{Fe}{2} nova for which we have good coverage.
Aside from near-IR observations \citep{N10}
and analysis of the super-soft X-ray source \citep{Swift11},
there has been little discussion of this bright nova.

\begin{figure}
\epsscale{0.8}
\includegraphics[scale=1.0,angle=0.]{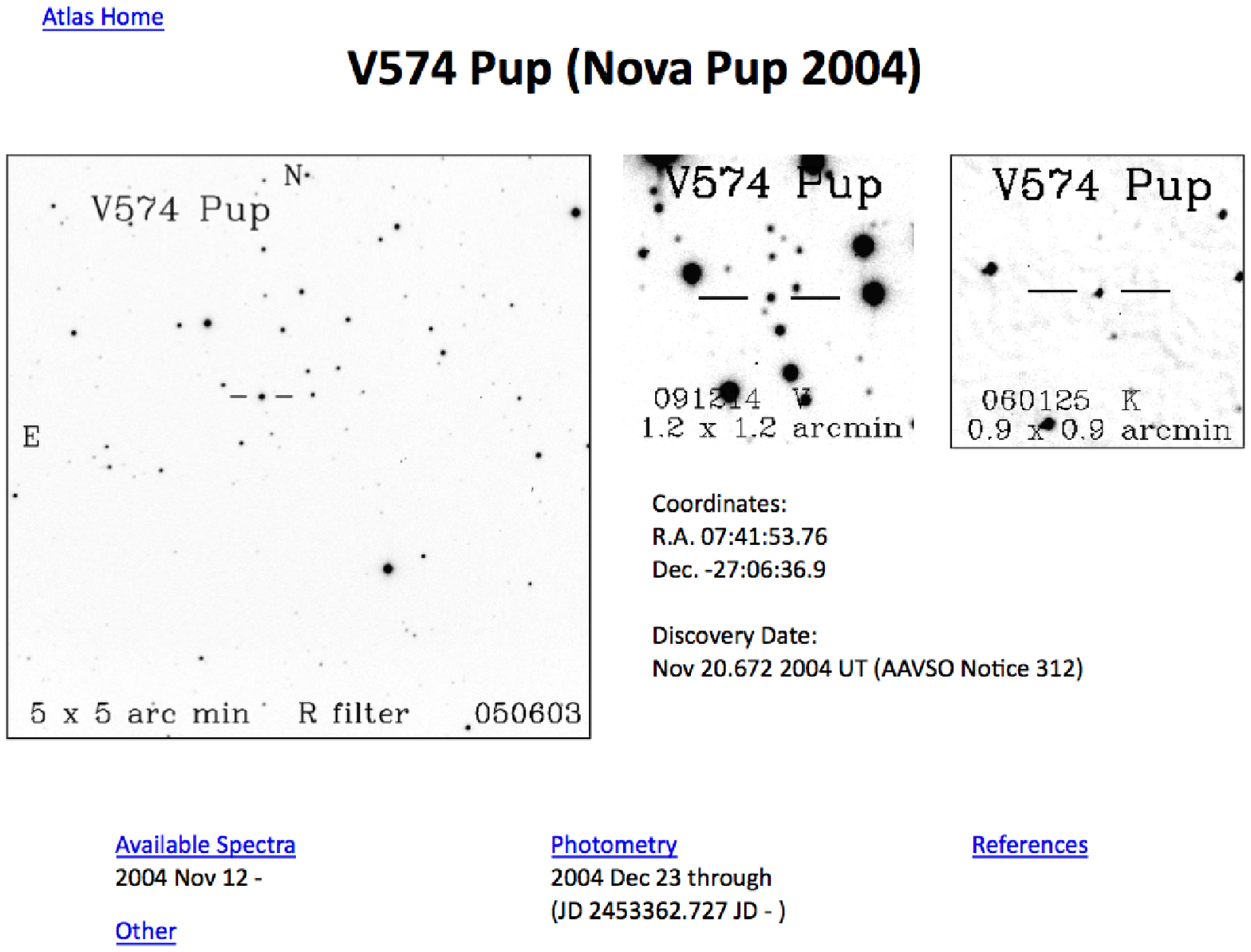}
\caption{The main atlas page for V574 Pup. The large finding chart 
shows the nova at an age of about 8 months. The smaller charts,
each $\sim$1~arcmin wide, show the nova in $V$ and $K$ after it has faded
substantially from peak.
\label{fv574main}}
\end{figure}

The main atlas page (Figure~\ref{fv574main})
presents finding charts in $V$ and $K$, along with the coordinates, time of
discovery, and links to the spectral and photometric data and references.

\begin{figure}
\epsscale{0.8}
\includegraphics[scale=.8,angle=-90.]{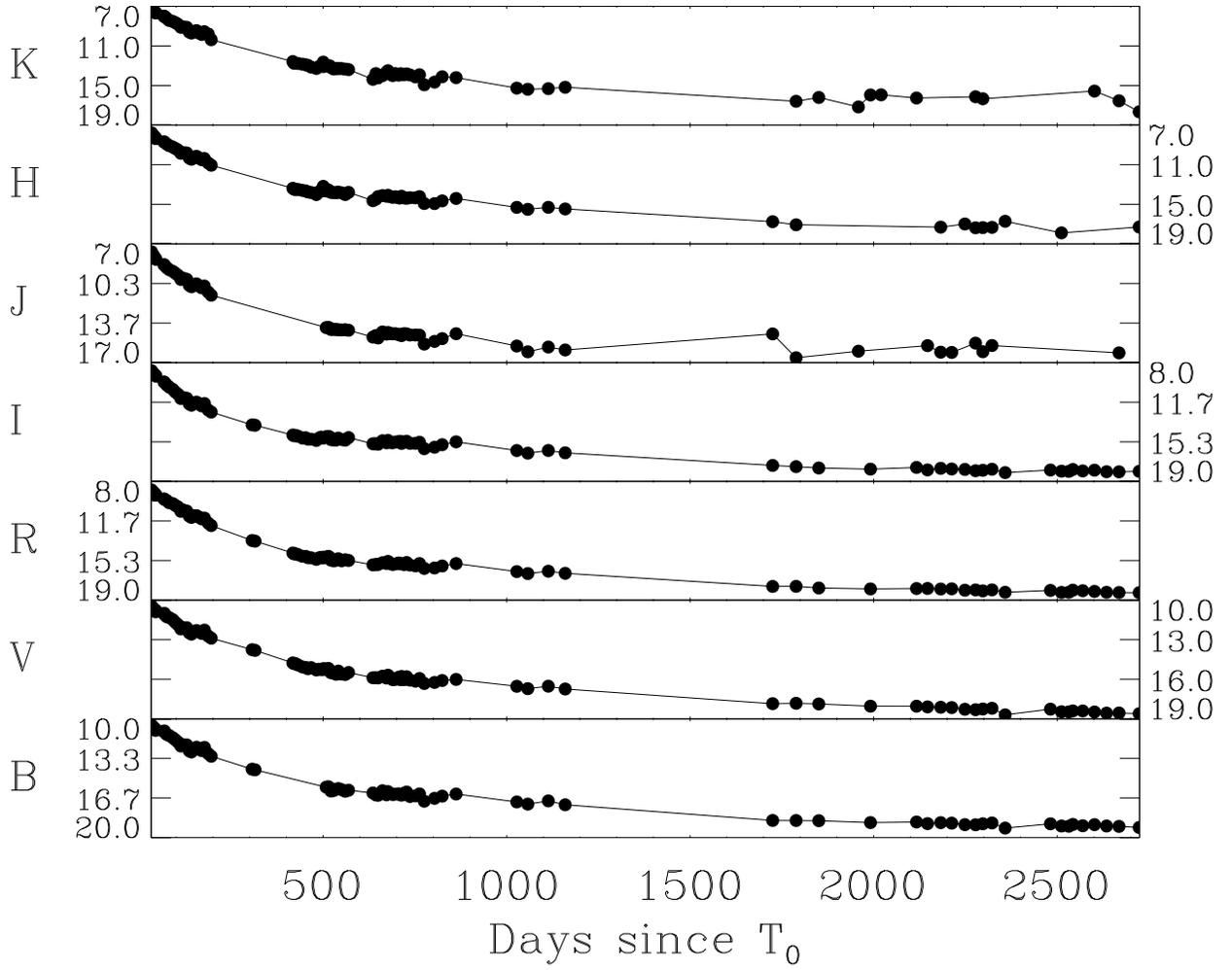}
\caption{SMARTS OIR light curve of V574~Pup from the ANDICAM
dual channel imager.
We observed on 100 days, starting on day 33 and running through day 2723.
The near-IR source is near the detection limit
after day 1500; only points with formal uncertainties $<$0.5~mag are plotted.
Error bars are smaller than the size of the points.
\label{v574_lc}}
\end{figure}

The photometry consists of observations taken on 100 days with the 1.3m
ANDICAM imager, starting on day 32 and
running through day 2723 (5 May 2012). Most of these sets include all 7 ANDICAM
bands, $BVRIJHK_s$. This light curve is shown in Figure~\ref{v574_lc}.
It is possible to fill in the first 30 days with data from other sources,
such as \cite{S05}, or by using data from the AAVSO (www.aavso.org).

\begin{figure}
\epsscale{0.8}
\includegraphics[scale=.8,angle=-90.]{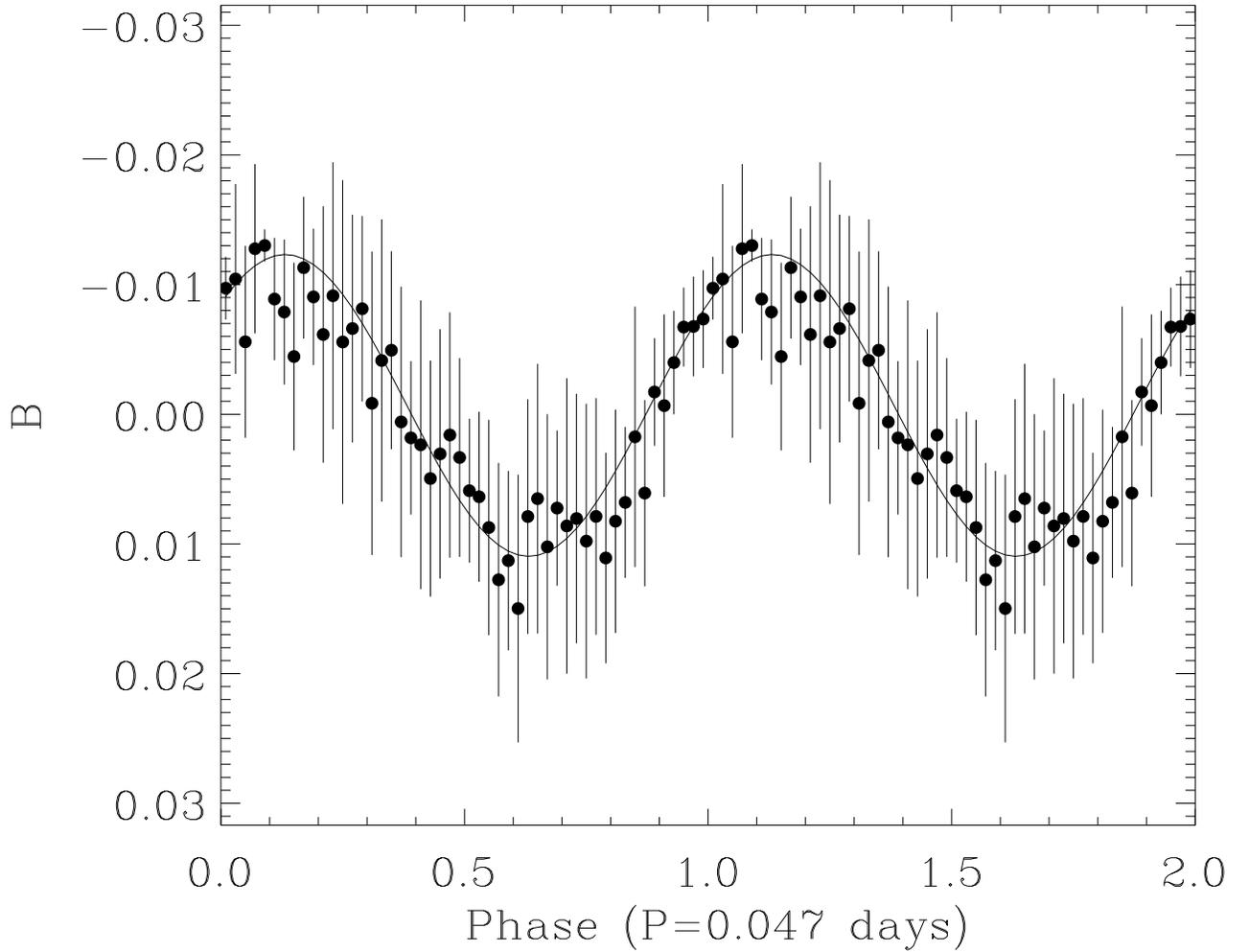}
\caption{366 $B$~band observations of V574~Pup made on days 87, 195, and 196
using the Apogee 0.5k CCD on the SMARTS 1.0m telescope. The data are folded 
on the best-fit period; two periods are shown. Phasing is arbitrary.
The 0.0472~day period is
determined by the shortest string method \citep{D83}; for the plot the data are
binned. Error bars represent the dispersion in each 0.02-cycle phase bin.
\label{fv574_var}}
\end{figure}

We supplemented these with data taken on 20 days using
a temporary small CCD on the SMARTS 1.0m telescope. These were opportunistic
observations enabled by the unavailability of the wide-field 4k camera.
We use the  $\sim$2 hour long sequences to search for short
periodicities. (Similar data exist for very few of the novae in the
atlas.) Three long sequences in the $B$~band,
on days 87, 195, and 196, showed sinusoidal-like modulations.
Removing a linear trend from the data on day 87 and normalizing to the mean
magnitudes, we find a likely period of 0.0472 days (68 minutes; see
Figure~\ref{fv574_var}) from a shortest-string analysis \citep{D83}.
However, we cannot exclude some aliases. This is shorter than the minimum
orbital period for CVs, and may be half an orbital period (ellipsoidal
variability is a possible explanation).
The amplitude of the best-fit sinusoid decreased from 0.02 to 0.007~mag
from day 87 to days 195-196.

We obtained 107 spectra before the target became too faint for the 1.5m
telescope.
We illustrate two types of investigations
that can be supported by high cadence spectral observations.

\begin{figure}
\epsscale{0.8}
\includegraphics[scale=.8,angle=-90.]{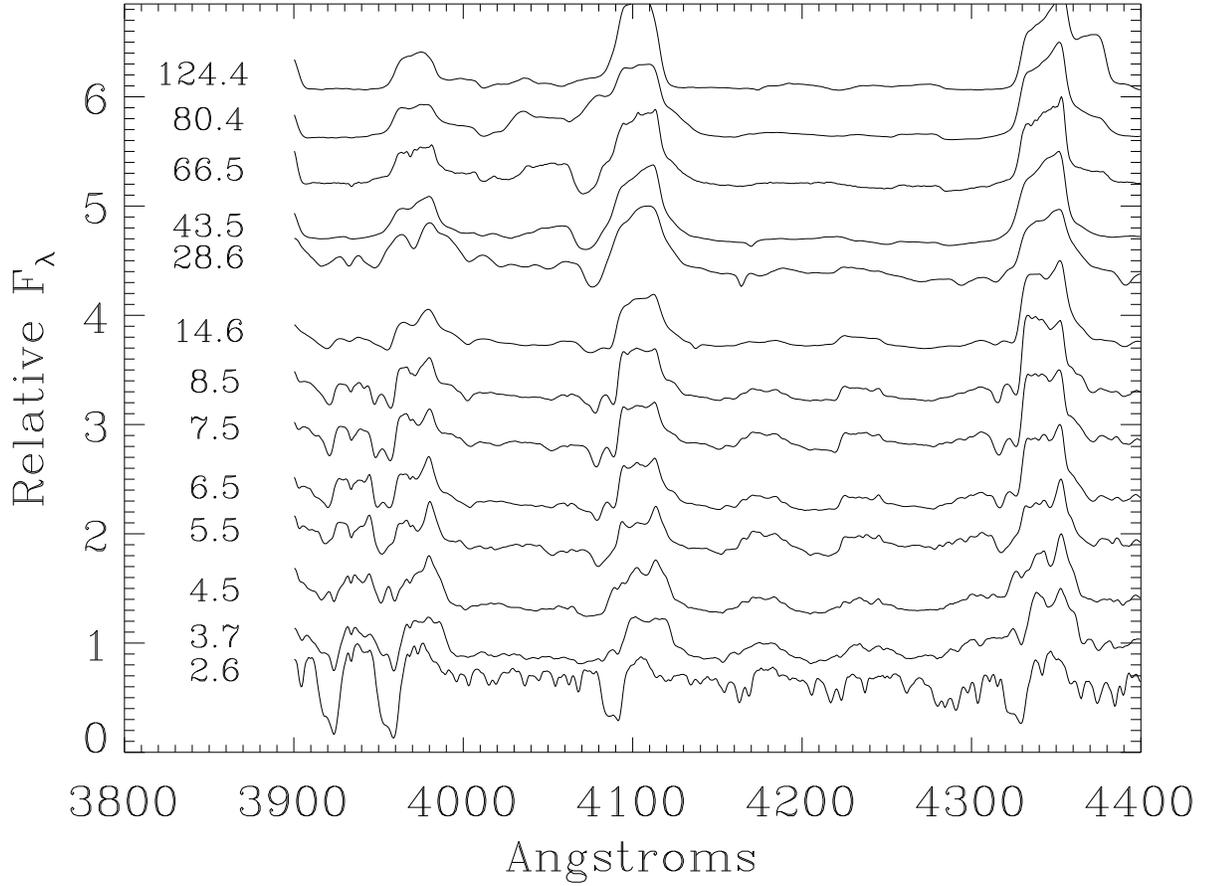}
\caption{The early spectral evolution of V574~Pup in the blue.
Spectra are normalized to the
peak brightness, and are offset by 0.5 units. The epoch in days is on the left
of the spectra. Resolution is 4.4\AA\ on days 66 and 80, and 1.6\AA\ otherwise.
The strong P~Cygni absorption in \ion{Ca}{2} K\&H and the
Balmer lines fades rapidly. The two distinct absorption features seen 
in the absorption blueward of H$\delta$ and H$\gamma$
seem to accelerate outward at different rates.
\label{fv574pc}}
\end{figure}

\begin{figure}
\epsscale{0.8}
\includegraphics[scale=.8,angle=-90.]{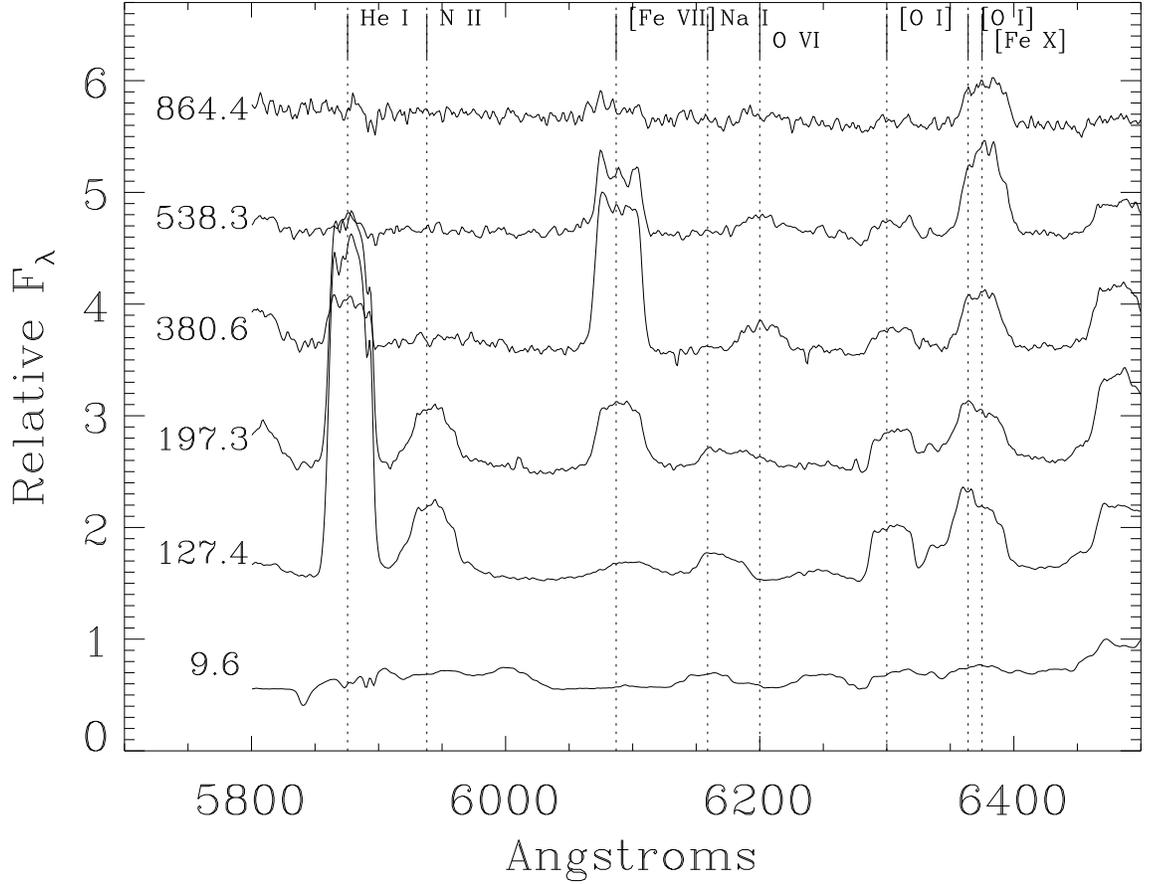}
\caption{The temperature evolution of V574~Pup from days 10 though 864.
Spectra are normalized to the peak brightness,
and are offset by 0.5 units. The epoch in days is on the left of the spectra.
Resolution is 3.1\AA.
The first spectrum (day 9.6) is dominated by continuum. Wind absorption from
\ion{He}{1} 5876 is seen, as are the interstellar sodium absorption lines.
By day 127 low excitation lines of \ion{He}{1}, \ion{N}{2}, and [\ion{O}{1}]
become important. The nebular [\ion{Fe}{7}] and [\ion{Fe}{10}] lines strengthen
later, peaking relative to the continuum after about 1 and 2 years respectively.
We show 6 of the 35 red spectra of V574~Pup, so a much finer time analysis is
possible.
\label{fv574fe}}
\end{figure}

\begin{enumerate}
\item Figure~\ref{fv574pc} shows the evolution of the P~Cygni line profiles
as the wind evolves against the backdrop of the optically-thick
pseudo-photosphere. With daily spectra, it is clear that the absorption
velocities are not constant, but rather are accelerating. Through day 14
the velocities can be described as a quadratic function of time. Hence the
acceleration is linear in time. It is hard to see how this can result
from decreasing optical depth effects in an envelope with a monotonic
velocity law increasing outwards. \cite{SNS11} show how similar structures
seen in T~Pyx can be explained as an outward-moving recombination front in
an envelope with a linear velocity law.


\item Figure~\ref{fv574fe} shows the time-evolution of a series of lines of
differing temperatures as the nova evolves through the nebular and coronal
phases. The [\ion{Fe}{10}] $\lambda$6375\AA\ line requires high excitation, and 
its presence correlates well with the super-soft (SSS)  X-ray emitting
phase \citep{Swift11}. V574~Pup was in its SSS phase from before
day 180 through day 1118; it ended before day 1312. We can use data such as
these to explore how well optical lines are diagnostic of the SSS phase.

\end{enumerate}

\subsection{Notes on the Novae}

Notes here are not meant to be complete or definitive in any sense. They are 
meant to highlight past or ongoing work on select novae, or to note some
particularly interesting cases. We have made no attempt to provide complete
references here; they are in the on-line atlas.
For the convenience of the reader, we have collected in
Table~\ref{tbl-meas} various basic measurements. These are:
\begin{itemize}
\item Spectroscopic class. This is a phenomenological classification
based on the appearance of the spectrum
in the first few spectra after the emission lines appear.
Physically, this is likely an
indicator of the optical depth of the envelope.
Of the 63 classifiable targets, most (47/63, or 75\%) are \ion{Fe}{2} type;
15 (24\%) are or may be He-N, and one is a possible symbiotic nova.
In one case we cannot tell because our first spectrum was obtained nearly  
2 years after peak.

We append a ``w'' in those cases where there is a clear P~Cygni absorption
in the Balmer lines (and sometimes in \ion{Fe}{2})
indicative of an optically-thick wind.
Half (32 of the 64 novae) show such P~Cyg absorption.
We caution that the absence of P~Cyg absorption may
be caused by the cadence of the observations.

\item Photometric class. We examined the $V$ band light
curves for the first 500 days and categorized them by eye into one or more
of the 7 classes defined by \cite{SSH10}. In many cases we have very little data
during the first 3 months, and do not attempt to categorize these. In some cases
we had a hard time shoe-horning the lightcurve into one class, and have given
multiple classes. For example, N LMC 2005 maintained a fairly flat light
curve for about 50 days (class F), then exhibited a cusp (class C). It also
formed dust (class D), though the dip is not particularly pronounced. The
presence of dust is indicated by the increase in the $H$ and $K$ fluxes
as the optical fades.

In many cases a significant brightening in $K$, suggestive of dust formation,
is not accompanied by an optical dip, suggesting an asphericity in the dust.

In some cases there is significant color evolution between the optical
and near-IR. We will quantify this later.

\item The FWHM of the H$\alpha$ emission line. We measure the first grating 47
spectrum (3.1\AA\ resolution) that does not show P~Cyg wind absorption,
and report
the day on which that spectrum was obtained. Uncertainties are of order 2\%.
Note that the FWHM can change significantly with time in the \ion{Fe}{2} novae.
For the He-N novae we measure the FWHM of the broad base, ignoring the narrow
central emission component. In some cases there is a faint but broader
component visible early on. The measurement of the FWZI of this component
would be more representative of the maximum expansion velocity.
We do not tabulate this because of incompleteness, and because of the
difficulty defining the continuum level in some cases.
\end{itemize}

We have not estimated the times for
the light to decay by 2 and 3 magnitudes at $V$
(t$_2$ and t$_3$, respectively) in
any systematic manner, because it is only in rare cases that we have 
sufficiently dense photometric sampling early enough to make a good estimate.
We discuss these in the notes on individual novae. 

We note that the estimates of t$_2$ and t$_3$ can be highly uncertain,
especially for fast novae. The reported discovery times are often past
the peak. The discovery magnitudes are often visual estimates, or 
unfiltered CCD magnitudes, necessitating a color correction to $V$.
A full analysis of the light curves,
incorporating other published literature, AAVSO data (which are much denser
near peak), and data from other
sources, is beyond the scope of this paper.

\subsubsection{N Aql 2005 = V1663 Aql}
This is a standard \ion{Fe}{2} nova. On day 50 there was prominent
$\lambda$4640\AA\ Bowen blend emission.
The auroral [\ion{O}{3}] lines were strong by day 85. 
Our last spectrum, on day 414, is dominated by
H$\alpha$, [\ion{O}{3}] 4959/5007, [\ion{N}{2}] 5755, [\ion{Fe}{7}] 6087,
[\ion{O}{1}] 6300, and [\ion{Ar}{3}] 7136.

\subsubsection{N Car 2008 = V679 Car}
This \ion{Fe}{2} nova never seemed to develop a coronal phase. 
We have limited photometric coverage.

\subsubsection{N Car 2012 = V834 Car}
This recent \ion{Fe}{2} nova exhibited a strong wind through day 36.
Evolution of the light curve has been uneventful. There was some jitter of
$\pm$0.5 mag from a smooth trend from days 12-40. We estimate
t$_2$ and t$_3$ to be 20 and 38 days, respectively, with uncertainties of order
$\pm$3 days for t$_2$ and $\pm$1 day for t$_3$.

\subsubsection{N Cen 2005 = V1047 Cen}
We have no photometry, and only two spectra, of this \ion{Fe}{2} nova.

\subsubsection{N Cen 2007 = V1065 Cen}
This dusty \ion{Fe}{2} nova was analyzed by \cite{Hel10}, using SMARTS spectra
through day 719. The atlas includes additional photometry, from days
944 though 1850.

\subsubsection{N Cen 2009 = V1213 Cen}
This \ion{Fe}{2} nova became a bright super-soft X-ray source. The coronal phase
extended from about days 300 to 1000, roughly coinciding with the SSS phase
\citep{Swift11}, with strong lines of
[\ion{Fe}{10}], [\ion{Fe}{11}], and [\ion{Fe}{14}].
In quiescence the remnant is blended with two other objects of
comparable brightness.

\subsubsection{PNV J13410800-5815470 = N Cen 2012}
This recent \ion{Fe}{2} nova exhibited wind absorption through day 25. 
t$_2$ is about 16$\pm$1 days; t$_3$ occurs about day 34.
The 2 mag brightening in $K$ starting about day 35,
with a contemporaneous drop in the $B$ and $V$~band brightness, 
suggests dust formation.
The strong emission in the \ion{Ca}{2} near-IR triplet on
day 11 had disappeared by day 74.

\subsubsection{PNV J14250600-5845360 = N Cen 2012b}
The $K$~band brightness increased by 2 magnitudes between days 18 and 32,
suggesting dust formation, but no drop is seen at optical magnitudes.
The smooth $V$ light curve yields t$_2$ and t$_3$ of 12.3 and 19.8 days, with 
uncertainties $<$1 day.
The spectral development is similar to N Cen 2012.
The strong emission in the \ion{Ca}{2} near-IR triplet on
day 16 had disappeared by day 61.

\subsubsection{N Cir 2003 = DE Cir}
This fast nova was discovered by \cite{L03} in the glare of the setting Sun.
Spectra obtained on days 11 and 12, at high air mass, show this was a He-N
nova. We did not obtain any photometry until after it reappeared from behind the
Sun. Since then it has been in quiescence at V$\sim$17, with a variance of
$\pm$0.4~mag. The strongest line in the quiescent spectrum is
He~II~$\lambda$4686\AA.

\subsubsection{N Cru 2003 = DZ Cru}
This is another nova that was discovered in the west in the dusk twilight.
Despite the discussion about the ``peculiar'' early spectrum \citep{iauc8185},
our spectra show this was an \ion{Fe}{2} nova discovered before maximum, as
concluded by \cite{Rus08}.

\subsubsection{N Dor 1937 = YY Dor}
This is the second recurrent nova discovered in the LMC \citep{lil04}. 
It is a fast (t$_2$,t$_3$=4.0, 10.9 days, respectively)
He-N nova with the broad tripartite Balmer lines seen in many
fast recurrent novae.
An analysis is in preparation.

\subsubsection{N Eri 2009 = KT Eri}
\begin{figure}
\epsscale{0.8}
\includegraphics[scale=.8,angle=-90.]{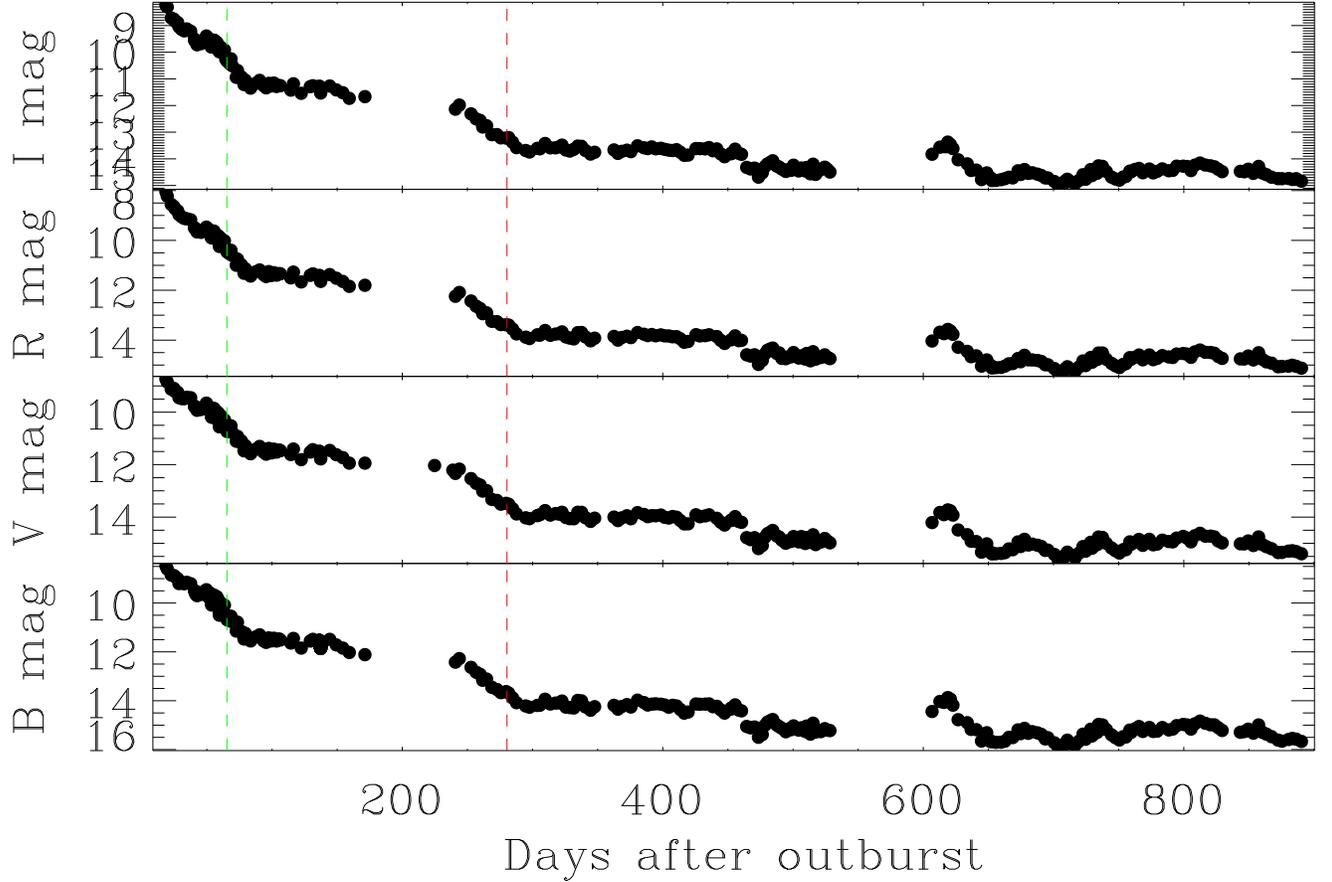}
\caption{The $BVRI$ light curves of KT~Eri from day 18 through day 890.
The two wide
gaps in the data are where KT Eri is too close to the Sun to be observed.
The initial
decay terminates about day 80, 2 weeks after the turn-on of the supersoft
X-ray source. Enhanced photometric scatter precedes the X-ray turn-on by
about 20 days. The first plateau runs from about days 80 through 210, followed
by a 2 magnitude fading over the next 3 months. The second plateau seems to
terminate with an abrupt 1 magnitude drop about day 460 to the level seen
in archival plates. Since then it has remained near $V$=15, with an RMS
scatter of 0.4 mag. The vertical dashed lines indicate the times of the
turn-on and turn-off of the bright super-soft X-ray source. 
\label{kteri-lc}}
\end{figure}

\begin{figure}
\epsscale{0.8}
\includegraphics[scale=1.0,angle=90.]{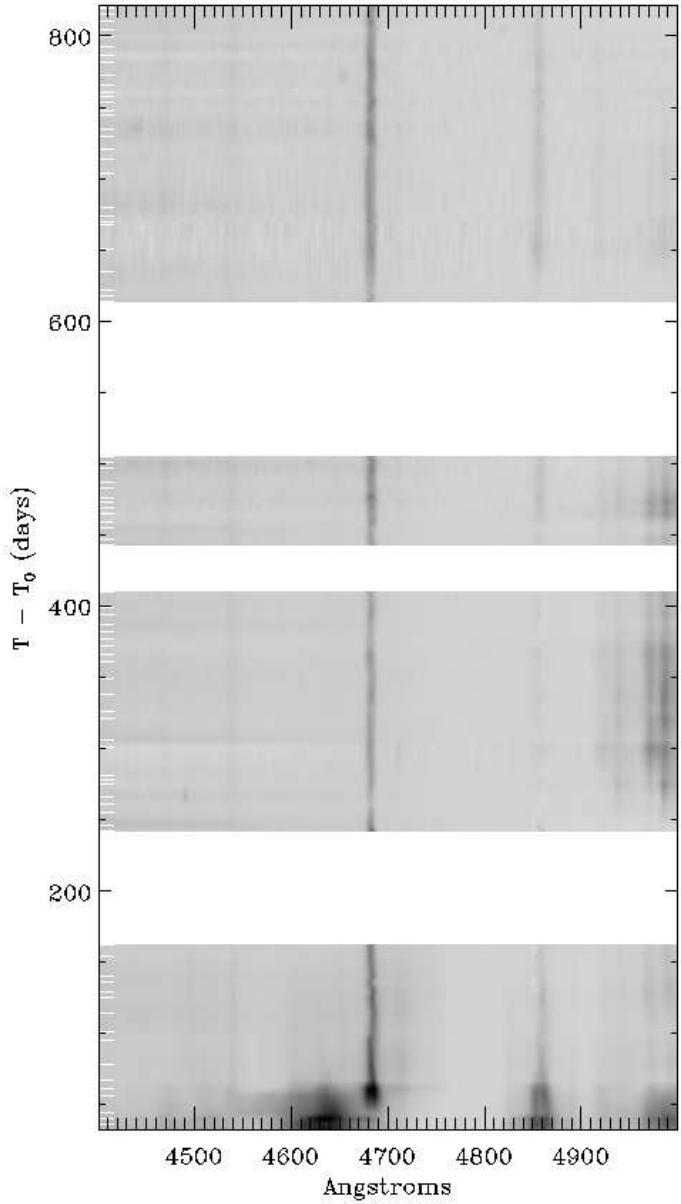}
\vspace{-1.0cm}
\caption{The trailed spectrum of KT Eri from day 32 through day 822,
constructed from 83 spectra (mode 26/Ia).
The times the spectra were obtained are indicated by the white ticks on
the left axis.
Data are linearly interpolated between spectra for gaps less than
30 days. The data are scaled to the 4750 -- 4820\AA\ continuum;
the intensity scaling is linear. During the first two
months the 4640\AA\ Bowen blend faded, \ion{He}{2} $\lambda$4686 turned on, 
and the  H$\beta$ line narrowed considerably, as did the [\ion{O}{3}]
$\lambda\lambda$4959,5007\AA\ lines. There is considerable intensity evolution
in the [\ion{O}{3}] lines. The late-time spectrum is dominated by \ion{He}{2}.
\label{kteri-trsp}}
\end{figure}

KT Eri is a fast He-N nova with a bright quiescent counterpart.
\cite{Hou10} reported a spectacular pre-maximum light curve from
the SMEI instrument. The light curve shows two plateaus (Figure~\ref{kteri-lc}), 
much like those seen in U~Sco, prior to dropping to quiescence.
\cite{Jur12} find a 737 day period in the quiescent source
from archival plate material. \cite{Hun11} claim a 56.7 day period 
during the second plateau. In quiescence, after day 650, 
there are hints of a period near 55 days,
and a possibly 4.2 day spectroscopic period (Walter et al. in preparation). 
Due to its brightness and RA (there is less competition for time towards the
galactic anti-center), we have excellent spectral time coverage.
Figure~\ref{kteri-trsp} shows the time-evolution of KT~Eri in the blue, over
790 days, from 83 low dispersion blue spectra.

KT~Eri is located in a sparse field; there is only a single comparison star
available in the small IR channel field of view.

\subsubsection{N Lup 2011 = PR Lup}
The light curve of this slow \ion{Fe}{2} nova showed
a second maximum about day 3.5. There was little appreciable decay
during the first 2 months.
Wind absorption was evident through day 58.
As of day 300 it has not entered the coronal phase. 

\subsubsection{N Mus 2008 = QY Mus}
We picked up this fairly late, but a combination of the
spectroscopy and photometry span the time that dust formed.
It seems to be a standard \ion{Fe}{2} nova that bypassed the coronal phase 
and is now in the nebular phase.

\subsubsection{N Nor 2005 = V382 Nor}
This appears to be a standard \ion{Fe}{2} nova.

\subsubsection{N Nor 2007 = V390 Nor}
We have good spectroscopic coverage of this \ion{Fe}{2} nova for 4  months,
but no photometry. During this time it did not evolve any hot lines.

\subsubsection{RS Oph}
This is the prototypical long period, wind-driven recurrent nova. 
The emission lines are narrow.
We are continuing observations to characterize it well into quiescence. 

\subsubsection{N Oph 2003 = V2573 Oph}
Based on two spectra, this appears to be a standard
\ion{Fe}{2} nova.

\subsubsection{N Oph 2004 = V2574 Oph}
Our photometric coverage consists of two observations, 4 and 8 years
after the eruption. Neither was obtained on a photometric night, so the
data are not yet photometrically calibrated.
We have good spectroscopic coverage showing the transition
from an optically thick pseudophotosphere to the permitted line spectrum in
this \ion{Fe}{2} nova.

\subsubsection{N Oph 2006 = V2575 Oph}
This is a standard \ion{Fe}{2} nova.

\subsubsection{N Oph 2006b = V2576 Oph}
Most photometric observations are in $R$ only. It is a standard
\ion{Fe}{2} nova.

\subsubsection{N Oph 2007 = V2615 Oph}
This is a normal \ion{Fe}{2} nova. Photometric coverage begins after 1.5 years.

\subsubsection{N Oph 2008 = V2670 Oph}
This is an \ion{Fe}{2} nova.

\subsubsection{N Oph 2008b = V2671 Oph}
This \ion{Fe}{2} nova faded rapidly, and was undetectable,
except at $R$, after 1 year.

\subsubsection{N Oph 2009 = V2672 Oph}
\cite{Mun11} reported on this very fast nova. t$_2$ and t$_3$ passed before our
first  photometry; \cite{Mun11} quote values of 2.3 and 4.2 days, respectively.
The spectra and spectral evolution are similar to those
of U~Sco. We followed this nova spectroscopically through day 31.6, and did not
see the deceleration reported by \cite{Mun11}.
This has the broadest H$\alpha$ line, at FWZI$\sim$11,000~km/s, of all the novae
in the atlas, exceeding that of U~Sco by about 20\%.

\subsubsection{PNV J17260708-2551454 = N Oph 2012}
This recent slow \ion{Fe}{2} nova showed no significant decline in
brightness from days 10 through 90. Then dust formed, with a drop in the
$B$ and $V$ brightness by over 5 magnitudes.
The lines are narrow: FWHM(H$\alpha$)$\sim$950~km/s.
There is little spectral evolution through day 90, with 
persistent P~Cygni line profiles.

\subsubsection{PNV J17395600-2447420 = N Oph 2012b}
This is an \ion{Fe}{2} nova with an  H$\alpha$ FWHM about 3000~km/s.
Through the first 45 days the brightness drops monotonically in $B$ through $K$.

\subsubsection{N Pup 2004 = V574 Pup}
See \S4.2. Our first photometry was on day 30, suggesting t$_3<27$ days. 
\cite{Swift11} quote t$_2$=13 days.

\subsubsection{N Pup 2007 = V597 Pup}
This slowly developing, broad-lined nova developed a coronal phase after
3-4 months.

\subsubsection{N Pup 2007b = V598 Pup}
This X-ray-discovered nova was in its nebular phase when reported by
\cite{RSE07}. The quiescent counterpart appears fairly bright.

\subsubsection{T Pyx}
This well-known recurrent novae is, as has been pointed out by many authors,
very different from the fast He-N recurrent novae. Its light curve and spectral
development mimic those of slow classical novae. 
The rate of the photometric decay remained unchanged by the turn-on/turn-off
of the SSS X-ray emission (Figure~\ref{tpyx-sss}).
The slope of the photometric decay decreased about day 300.
The H$\alpha$ line width shown in Table~\ref{tbl-meas} refers to the
width of the broad base after the line profile stabilized; it was about half
that during the first 40 days post-peak.

\cite{Eva12} include some of our near-IR photometry in an analysis of the
heating of the dust already present in the system.

As T Pyx remains bright, we are continuing our monitoring.

\subsubsection{U Sco}
This is the prototypical short orbital period recurrent nova.
Some spectral analysis in included in \cite{Max12}.

\subsubsection{N Sco 2004b = V1187 Sco}
This well-observed \ion{Fe}{2} nova became a super-soft X-ray source.

\subsubsection{N Sco 2005 = V1188 Sco}
We have only limited coverage of this \ion{Fe}{2} nova.

\subsubsection{N Sco 2007a = V1280 Sco}
This extraordinarily slow nova has remained bright ($V$ generally
between 10 and 11) for nearly 2000 days. \cite{N12} present a lot of data
for this nova; our monitoring provides finer temporal coverage, which
show absorption events with durations $<$60 days that may be due to dust
formation in small mass ejection events. The latest spectra, at an age of 5.3
years, still show evidence for wind absorption.

\subsubsection{N Sco 2008 = V1309 Sco}
This is a very narrow-lined system, and likely a symbiotic nova
or a merger \citep{Mas10}. The spectrum is dominated by narrow Balmer line
emission.

\subsubsection{N Sco 2010 No.~2 = V1311 Sco}
This nova faded rapidly. It is likely a He-N class nova. On day 9
possible [Ne III] $\lambda$3869 is seen, and the $\lambda$4640 Bowen blend
is in emission.

\subsubsection{N Sco 2011 = V1312 Sco}
This \ion{Fe}{2} nova developed coronal line emission.

\subsubsection{N Sco 2011 No.~2 = V1313 Sco}
This nova exhibited very strong \ion{He}{1} and H-Paschen line emission.
Strong \ion{He}{2}
$\lambda$4686 emission appeared between days 16 and 19.
The Balmer lines have a narrow central core atop a broad base, as in the He-N
and recurrent novae, but there is also likely \ion{Fe}{2} multiplet 42
emission at
$\lambda$5169\AA\ (other multiplet 42 lines are overwhelmed by \ion{He}{1}
emission lines), and wind absorption at least through day 3. This seems to be
a hybrid nova.
t$_2$ lies between 5.8 and 8.5 days, and t$_3$ between 13 and 18 days,
depending on whether the peak was on
2011 Sep 6.37 or 2011 Sep 7.51 \citep{Sea11}.
The continuum is red. This may be a symbiotic nova.

\subsubsection{N Sct 2003 = V475 Sct}
This was the first nova that we concentrated on. Results appear in 
\cite{Str06}. This fairly narrow-lined \ion{Fe}{2} nova may have formed dust;
no coronal phase was seen. We have $U$-band photometry for the first 100 days,

\subsubsection{N Sct 2005 = V476 Sct}
We have very limited observations of this \ion{Fe}{2} nova.

\subsubsection{N Sct 2005b = V477 Sct}
This is probably an \ion{Fe}{2} nova; we have very poor coverage.

\subsubsection{N Sct 2009 = V496 Sct}
This \ion{Fe}{2} nova likely formed dust. As it is fairly bright, we have good
spectral coverage for nearly 2 years.

\subsubsection{N Sgr 2002c = V4743 Sgr}
This is the first nova we started observing. We picked it up about 200 days
after discovery. There is currently no photometry in the atlas.

\subsubsection{N Sgr 2003 = V4745 Sgr}
This was the first nova to explode during the SMARTS era.
There are two prominent P~Cygni absorption line systems,
initially at -780 and -1740
km/s, visible from days 12 through 66; they disappear by day 71.
There is currently no photometry in the atlas.

\subsubsection{N Sgr 2004 = V5114 Sgr}
Data for this \ion{Fe}{2} nova have been analyzed and published by \cite{E06}.
\cite{E06} quote t$_2$ and t$_3$ values of 11 and 21 days; for a peak
$V$=8.38 we find a marginally slower nova, with t$_2$ and t$_3$ of 14 and
25 days, respectively.
We have $U$ band photometry for the first 180 days.

\subsubsection{N Sgr 2006 = V5117 Sgr}
This is a standard \ion{Fe}{2} nova. We estimate t$_2$ and t$_3$ are about
16$\pm$1  and 42 days, respectively, with uncertainties of perhaps a week in t$_3$.

\subsubsection{N Sgr 2007 = V5558 Sgr}
This very slow nova resembles V723~Cas. We started the photometry at about
day 100; the nova remained at $7.0<V<9$ through day 200,
with the exception of one
short dip to $V$=10 seen in all 7 bands. Since then there has been an
uneventful decay to $V\sim14$. It is a narrow-lined \ion{Fe}{2} nova that 
exhibits P~Cygni absorption through day 208. \ion{He}{2}~4686 exceeded
H$\gamma$ in strength by day 480, and rivaled H$\beta$ by day 1100. The
[\ion{Ne}{5}] doublet was visible by day 432, and became the strongest line,
aside from H$\alpha$, by day 1150. There is very strong [\ion{Fe}{7}] emission,
but little coronal [\ion{Fe}{10}].

\subsubsection{N Sgr 2008 = V5579 Sgr}
This standard \ion{Fe}{2} nova apparently formed dust, because it was bright
in the near-IR ($K$=6.6) and fainter than 23$^{rd}$~mag at $BVRI$ on day 68.
When we next looked at it, on day 1120, V$\sim16.1$~mag with $V-K\sim$1.6.

\subsubsection{N Sgr 2009 No.~3 = V5583 Sgr}
This appears to be a standard \ion{Fe}{2} nova. Linear interpolation between our
first 2 observations yields t$_2$ and t$_3$ = 6.7 and 12.6 days, respectively;
\cite{Swift11} quotes t$_2\sim$5 days.

\subsubsection{N Sgr 2009 No.~4 = V5584 Sgr}
This appears to be a standard \ion{Fe}{2} nova.

\subsubsection{N Sgr 2010 No.~2 = V5586 Sgr}
We have only a single red spectrum of this nova. The broad H$\alpha$ emission
line, the presence of strong broad \ion{He}{1} lines, and the rapid decay 
are all consistent with a He-N classification. The nova is located about
2.5~arcsec E of a highly extincted near-IR source, 2MASS J17530316-2812183.
This can contaminate the photometry, particularly on nights with less
than ideal seeing.

\subsubsection{N Sgr 2011 No.~2 = V5588 Sgr}
This \ion{Fe}{2} nova exhibited at least 6 outbursts of up to 2~mag
during the first 200 days. 
The spectrum shows an \ion{Fe}{2} nova with very strong and
time-variable \ion{He}{2}~4686 and [\ion{Fe}{7}] and [\ion{Fe}{10}]
emission lines.

\subsubsection{PNV J17452791-2305213 = N Sgr 2012}
This fast hybrid nova
showed \ion{Fe}{2} emission on day 3, but by day 8 resembled
a He-N nova. By day 65 it was in its coronal phase, with emission from
[\ion{Fe}{10}], [\ion{Fe}{11}], and [\ion{Fe}{14}].
The FWHM of H$\alpha$ is about 5700~km. t$_2$ passed prior to our first
photometric observation, and was likely 4.5$\pm$1.5 days; t$_3$ is about 7 days.

\subsubsection{N TrA 2008 = NR TrA}
\begin{figure}
\epsscale{0.8}
\includegraphics[scale=.8,angle=-90.]{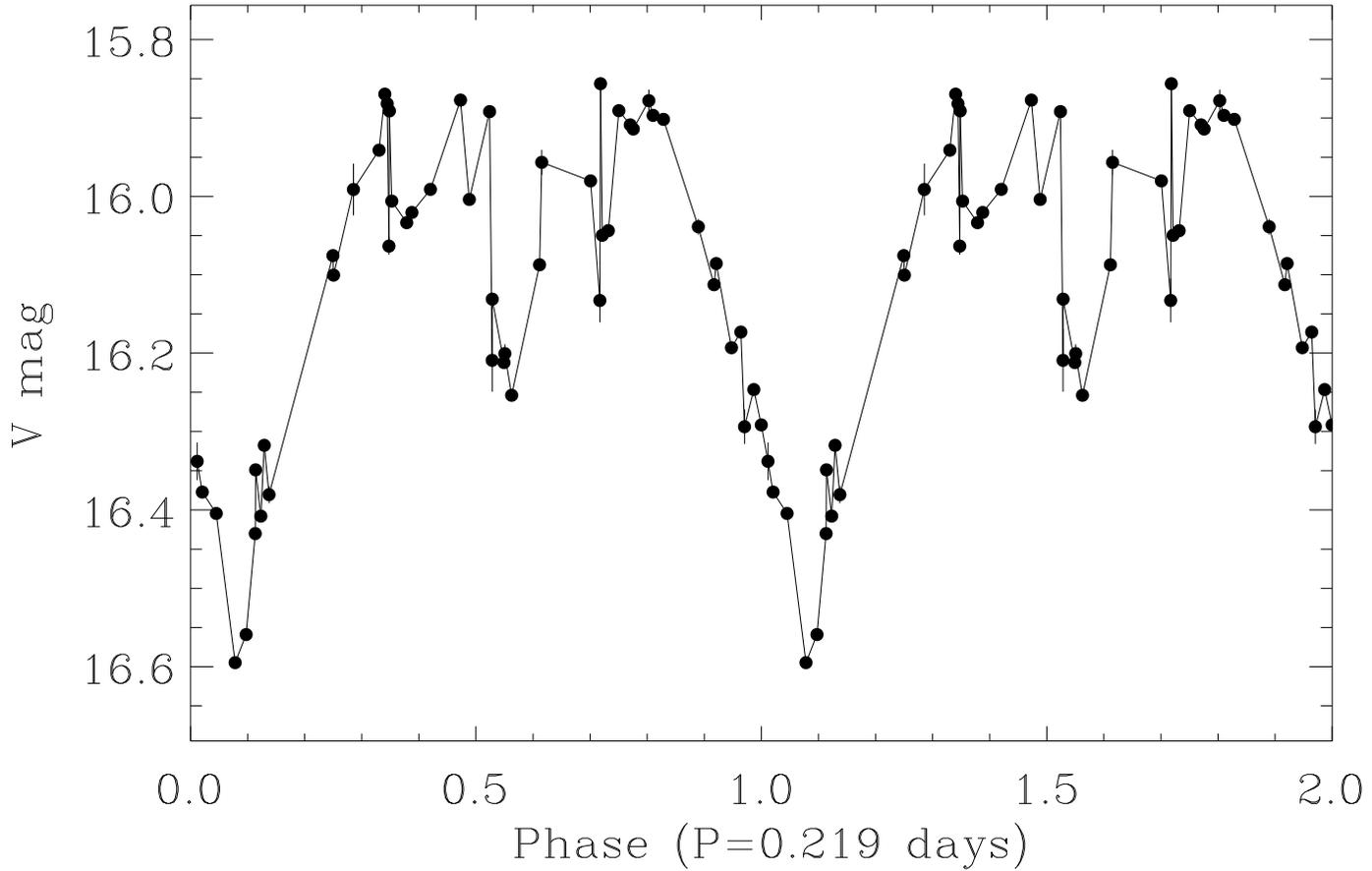}
\caption{The $V$~band light curve of NR TrA from days 1398 through 1534
after peak, folded on the 5.25 hour period. Two cycles are plotted. 
The broad dip covers nearly 0.5 cycles, is about 0.7 mag deep,
and is triangular in shape. There may be a secondary 
dip in anti-phase, but there also appears to be variability at the
0.1~mag level at all phases. The $BRI$ light curves are similar.
51 observations make up this light curve.
\label{nrtra-vlc}}
\end{figure}

This is the only nova in our collection that has shown eclipses.
The 5.25 hour period has a broad primary minimum covering half the period
(Figure~\ref{nrtra-vlc}) and a likely smaller secondary minimum.

\begin{figure}
\epsscale{0.8}
\includegraphics[scale=.8,angle=-90.]{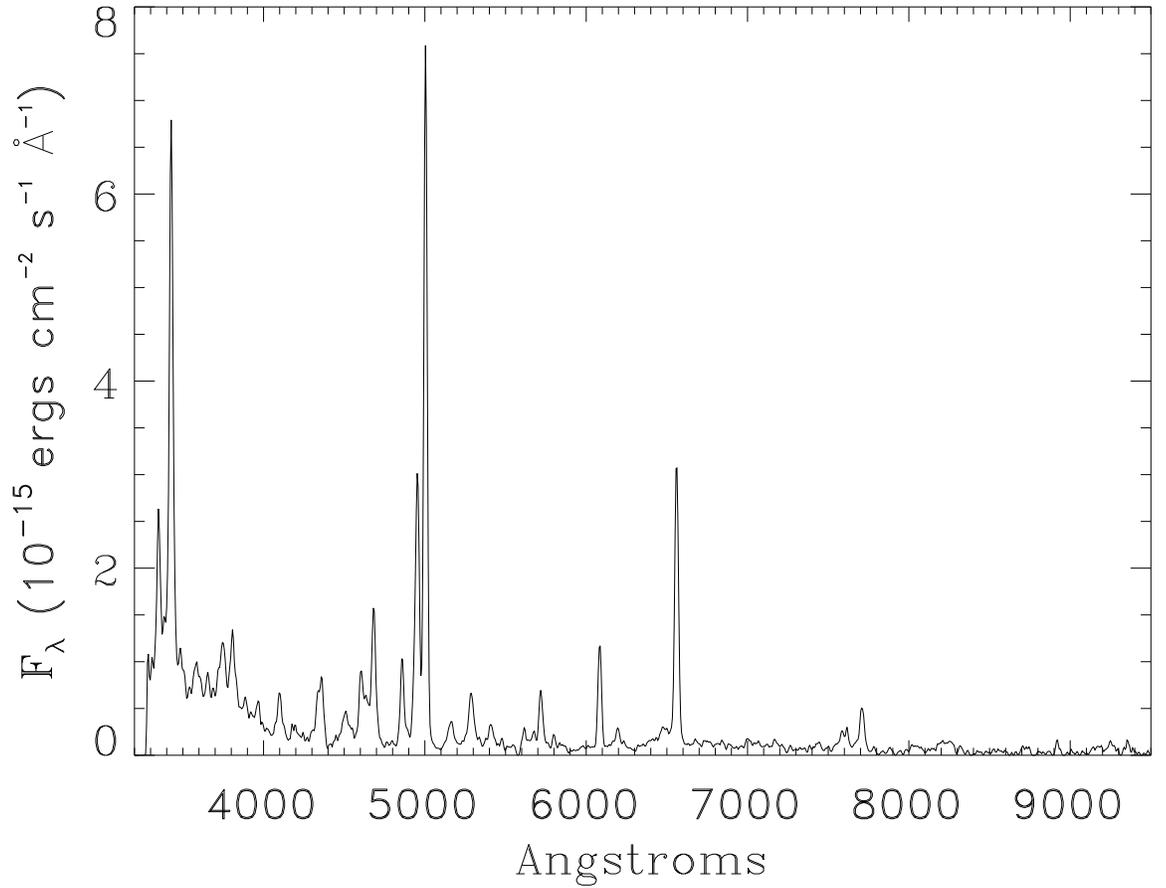}
\caption{The full optical spectra (13/I mode) of NR~TrA on 15 March 2012
(day 1444). The nova is classified as nebular \citep{Wil91}
because the strongest
non-Balmer line is [\ion{O}{3}] $\lambda$5007. Aside from the nebular lines,
strong permitted lines of \ion{He}{2}, \ion{N}{3}, \ion{N}{4} or \ion{N}{5},
and \ion{O}{6} are visible.
\label{nrtra-optsp}}
\end{figure}

\begin{figure}
\epsscale{0.8}
\includegraphics[scale=.8,angle=-90.]{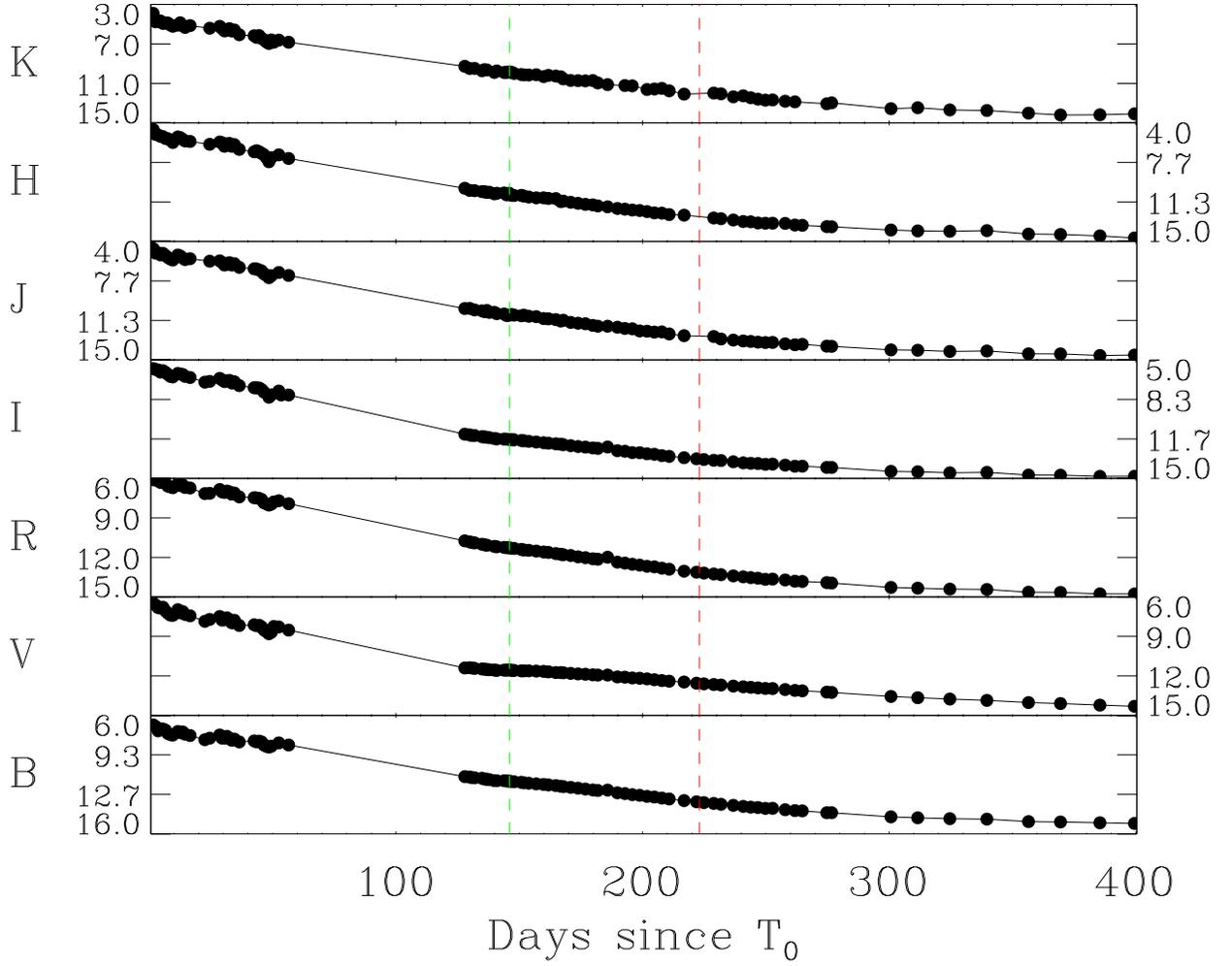}
\caption{The post-peak OIR light curve of T Pyx. The dashed lines represent the
approximate times of the turn-on and turn-off of the super-soft X-ray emission.
Note that there is neither a plateau in the light curve nor any change in the
slope of the decay near these times. The rate of decay did slow between days
250 and 300. 
\label{tpyx-sss}}
\end{figure}

The early spectra are those of an \ion{Fe}{2} nova with narrow lines.
The cool permitted lines (\ion{N}{2}, \ion{He}{1}, \ion{Fe}{2})
faded out between days 570 and 1220, and were replaced with a composite of a
nebular spectrum with a WN-like spectrum.
After the nebular lines of [\ion{O}{3}], [\ion{Ne}{5}], and [\ion{Fe}{7}],
the strongest lines are the Balmer series and \ion{He}{2} $\lambda$4686. 
Other prominent high excitation lines include
\ion{He}{2} (4200, 5411, 7592, 8236)
\ion{N}{3} (4515 and the 4640 blend),
\ion{N}{4} and/or \ion{N}{5} (4606/4608),
\ion{C}{4} (5803),
\ion{O}{6} (3811, 5292, 6200),
and a strong line that may be a blend of \ion{N}{4} (7703), \ion{C}{4} (7708),
and \ion{O}{4} (7713).
A recent optical spectrum is shown in Figure~\ref{nrtra-optsp}.

Aside from the nebular lines, the system bears a striking resemblance to
the V~Sge stars \citep{SD98}. \cite{SD98} argue that the presence of \ion{O}{6},
and the absence of strong hard X-ray emission, implies the presence of a
source of super-soft X-rays, with kT between 30 and 50~eV. This in turn 
suggests steady nuclear burning in the atmosphere of a massive WD, as in
the persistent SSS sources like Cal~83 or RX~J0513-69. However, models
containing other types of compact objects, or even WC stars, remain viable,
and \cite{SBZ01} argue that V~Sge is a contact binary following common
envelope evolution. However, \cite{HK03} conclude that V~Sge is the product of
accretion wind-driven evolution of a massive WD, similar to the persistent
SSS sources.

Since NR~TrA, as a classical nova, must contain a WD, if further investigation
confirms its resemblance to the V~Sge stars then we may be able to clarify the
nature of those systems.

\subsubsection{PNV18110375-2717276 = N Sgr 2012b}
This apparently very slow nova was reported in April 2012 
\citep{Nak12}, and then found to have been bright at least as early as 12 April
2011.

\subsubsection{PNV17522579-2126215 = N Sgr 2012c}
This nova was reported on 26 June 2012. 
The single spectrum we have shows a broad H$\alpha$ line, with FWHM~4200~km/s.
Both t$_2$ and t$_3$ passed prior to our first photometric observations, with
t$_3<$7.5 days.

\subsubsection{N LMC 2005}
This is a slow \ion{Fe}{2} nova. The light curve was flat for about 2 months,
developed a cusp, and then formed dust after about 100 days.

\subsubsection{N LMC 2009}
This is a recurrence of N LMC 1971b, and becomes the third known
recurrent nova seen in the LMC. It developed into a strong
super-soft X-ray source.
A full analysis is in preparation (Bode et al. 2012).

\subsubsection{N LMC 2009b}
This \ion{Fe}{2} nova formed dust; there is a very deep trough in the
optical light curve between about days 80 and 120, following a 4~mag increase
in brightness at $K$.

\subsubsection{N LMC 2012}
This is a very fast (t$_2$=1.1$\pm$0.5 days; t$_3$=2.1$\pm$0.5 days)
recent He-N nova.

\subsubsection{CSS081007:050559+054715}
This peculiar high galactic latitude object showed triple peaked,
velocity-variable H$\alpha$ emission. After [\ion{Ne}{5}], the strongest
emission line is \ion{He}{2}~$\lambda$4686\AA.

\subsubsection{XMMU J115113.3-623730}
The spectra nearly 2 years after peak shows strong emission in \ion{He}{2},
\ion{C}{4}, \ion{N}{4}, and \ion{O}{6} (see \citealt{Hug10}).
We confirm the 8.6 hour
period \citep{Pat10}. The system bears some resemblance to
high excitation systems like V~Sge. The nebular [\ion{O}{3}] emission, absent on
day 604, appeared about day 800 and now dominates the spectrum (day 1281).


\section{Discussion}\label{sec-disc}
The nature of our interest in novae has evolved over time, as we have gained
experience with these systems. What seemed at first to be a largely understood
class of objects has become more and more puzzling as the full scope of the
population has become evident.  

We are exploiting this database to investigate a number of areas, including:
\begin{itemize}
\item The origin and evolution of the tripartite Balmer line profiles in the
   fast He-N novae. 
   Such novae eject very little mass, hence their envelopes are thin, and
   afford the opportunity to view the
   innermost environs of the novae at early times.
   \cite{WB11} note the similarity of the tripartite Balmer line profiles
   to those of optically-thin accretion disks, and suggest that we are
   seeing either a disk that survives the outburst, or one that reconstitutes
   itself within a few days of the outburst.

\item The relation of the spectral and photometric evolution to the super-soft
   X-ray emission. The X-rays probe the innermost, hottest regions near the
   surface of the white dwarf. While the envelope is optically thick to
   soft X-rays, one can probe hotter regions using the Bowen fluorescence
   mechanism \citep{MCT75, KM80} and the optical \ion{He}{2} lines. We are
   examining how these lines vary with the emergent soft X-ray flux.

\item Novae in the LMC. These have the advantage over novae in the Milky
Way in that they are at a known distance, and suffer fairly small reddening. 
They are a small but statistically complete sample that can be used for
population studies. A catalog of all known novae in the LMC is maintained at
MPE\footnote{\url{www.mpe.mpg.de/\~{}m31novae/opt/lmc/index.php}}. 
While a much larger
sample of novae at a uniform distance and low reddening exists in M31
(e.g., \citealt{Pie07}, \citealt{Sha11}),
those novae are fainter and the field is more crowded.
The novae in the LMC can generally be followed for longer and in more detail
than can those in M31.

\end{itemize}

Novae are complex systems that likely lack spherical symmetry, hence
examination of the evolution of a large sample will provide insights into
the behaviors that may be masked by geometric effects in particular cases.
Our investigations focus on the ensemble for this reason, but there are many
other types of investigations that these data can help one address.

\section{Access to the Data}\label{sec-access}
We are making the data freely available to the community at
\begin{center} 
http://www.astro.sunysb.edu/fwalter/SMARTS/NovaAtlas/
\end{center}
with the following caveats:
\begin{itemize}
\item There may be faulty data in the atlas. This includes some early low
      dispersion spectra (grating 13 setup) that are clearly mis-calibrated. 
      We will correct this as time permits. 
\item There are some spectra that are clearly not of the correct star.
      We have tried to catch these, but some certainly have escaped notice.
\item There are mis-calibrated or poorly calibrated spectra for the
      reasons alluded to in \S\ref{sec-lds}.
\item Magnitudes of very faint novae, or novae in crowded regions, should be
      used with caution. At a later time we may employ PSF-fitting photometry.
\item Magnitudes are differential, but corrected for zero-point, and may differ
      systematically from the truth.
\item Some data are not posted, simply because we are currently working on
      those objects.
      
\end{itemize}

If you choose to use any of the data in your research, please reference this
paper. Any questions about the data may be directed to the lead author.

\acknowledgments

The compilation of this atlas 
would not have been possible had it not been for the
vision of Charles Bailyn, who formed the SMARTS partnership in 2003 in order
to keep the small telescopes at CTIO,
and the science enabled by them, open
and accessible to the community. 

The reason that SMARTS is a success is due in large part to its cadre of
service observers, including Claudio Aguilera, Sergio Fernandez,
Rodrigo Hernandez, Manual Hernandez, Alberto Miranda,
Alberto Pasten, Jacqueline Seron, and Jose Velasquez. They have taken the
vast majority of the observations available in the atlas.
Their professionalism ensures the uniformly excellent quality of the data.
Eduardo Cosgrove and Arturo Gomez have been invaluable in keeping the
telescopes and instruments running.
We are thankful for the efforts of the 1.3m telescope schedulers at Yale,
including
M. Buxton, R. Chatterjee, and J. Nelan, to accommodate our many requests for
prompt scheduling of new novae.

We thank W. Liller for forwarding reports of his discoveries.

FMW is desperately trying to learn enough to expound confidently about novae,
and is grateful to B. Schaefer, G. Schwarz, S. Shore, R. Williams,
and other members of the {\it Swift} Nova working group for
enlightening discussions.

We acknowledge support from the Provost, the Vice President for Research,
and the Department of Physics \& Astronomy of Stony Brook University
to purchase time at SMARTS. We acknowledge support from HST GO grants to
Stony Brook University to obtain ground-based observations in concert with
HST UV spectroscopy. HEB acknowledges support from the STScI Director's
Discretionary Research Fund, which supported STScI's participation in the
SMARTS consortium.

\clearpage

\begin{deluxetable}{lrrr}
\tablewidth{0pt}
\tablecaption{Spectrograph Setups \label{tbl-spsetups}}
\tablehead{
\colhead{Mode}  & \colhead{Resolution}   & \colhead{Filter} & \colhead{Wavelength Range}\\ & \colhead{(\AA)} & & \colhead{(\AA)}}
\startdata
\multicolumn{4}{c}{\underline{Standard Modes}}\\

13/I   & 17.2 & clear & 3146$-$9374\\
26/Ia  &  4.1 & clear & 3660$-$5440\\
47/IIb &  1.6 & BG39  & 4070$-$4744\\
47/Ib  &  3.1 & GG495 & 5650$-$6970\\

\multicolumn{4}{c}{\underline{Other Modes}}\\
9/I    &  8.6 & clear    & 3500$-$6950\\
47/II  &  1.6 & CuSO$_4$ & 3878$-$4552\\
58/I   &  6.5 &	GG495    & 6000$-$9000\\
\enddata
\end{deluxetable}

\small
\begin{deluxetable}{lll|rrrr|rrr}
\tablewidth{0pt}
\tablecaption{Observational Details - ``Old'' Novae\label{tbl-obs}}

\tablehead{\multicolumn{2}{c}{Nova} & \multicolumn{1}{c}{T$_0$\tablenotemark{a}}
 &\multicolumn{4}{c}{Photometry} & \multicolumn{3}{c}{Spectroscopy}  \\
\colhead{}  & \colhead{}  & \colhead{}  &
\colhead{Nights}  & \colhead{Start}  & \colhead{End} & \colhead{$V$} &
\colhead{Nights}  & \colhead{Start}  & \colhead{End}  } 
\startdata
N Aql 2005 & V1663 Aql   &3530.7& 32 & 12.0 & 2264.9 & $>$22 & 16 & 48.9 &  414.
0 \\
N Car 2008 & V679 Car    &4797.8& 39 & 75.0 & 1255.8 &  18.5 & 51 &  3.1 &  606.
7 \\
N Cen 2005 & V1047 Cen   &3614.5&  0 & ---  & ---    &  ---  &  2 &  4.9 &    6.
9 \\
N Cen 2007 & V1065 Cen   &4123.9&  9 & 14.9 & 1850.9 &  18.1 & 52 &  5.9 &  715.
9 \\
N Cen 2009 & V1213 Cen   &4959.7& 55 & 88.8 & 1122.8 & $>$17 & 31 &118.9 & 1016.
9 \\  
N Cir 2003 & DE Cir      &2921.5& 99 &161.2 & 3173.1 & 17.2  & 62 & 11.0 & 3074.
2 \\
N Cru 2003 & DZ Cru      &2871.9&  0 & ---  & ---    & ---   &  6 &  1.6 &   21.6 \\
N Eri 2009 & KT Eri      &5150.1&265 & 18.5 &  890.3 & 15.4  &368 & 12.5 &  845.
3  \\
N Lup 2011 & PR Lup      &5784.& 40 &  1.6 &  310.6 & 15.0  & 36 & -1.5 &  299.6
  \\
N Mus 2008 & QY Mus      &4738.5& 67 &134.2 & 1317.1 & 16.3  & 40 & 86.3 & 1224.
3  \\
N Nor 2005 & V382 Nor    &3442.8& 19 &307.1 & 2556.0 & 17.6  & 14 & 11.9 &  521.
8  \\
N Nor 2007 & V390 Nor    &4268.2&  0 & ---  & ---    & ---   & 34 &  4.6 &  128.
3  \\
N Oph 1898 & RS Oph      &3779.3& 57 & 15.5 & 2303.5 & 11.6  & 87 & 23.6 & 2022.
3  \\
N Oph 2003 & V2573 Oph   &2018.3&  0 & ---  & ---    & ---   &  2 &822.4&823.4\\
N Oph 2004 & V2574 Oph   &3110.3&  2 &1573.4& 2901.5 & ---   & 50 &  1.4 &  851.
4  \\ 
N Oph 2006 & V2575 Oph   &3775.9& 17 & 16.0 & 1135.9 & $>$22 &  5 & 27.0 &  869.
7  \\
N Oph 2006b & V2576 Oph  &3832.1& 16 &  9.7 & 2179.7 & 19.7(R)&17 & 32.7 &  191.
4  \\
N Oph 2007 & V2615 Oph   &4179.3&  6 & 18.2 & 1619.3 & 21.0  & 13 & 11.5 &  149.
3  \\ 
N Oph 2008 & V2670 Oph   &4612.5& 13 & 31.3 & 1445.2 & 20.8  & 18 & 22.3 &  480.
0  \\
N Oph 2008b & V2671 Oph  &4618.1&  9 & 25.7 & 1389.8 & $>$22 & 14 & 16.7 &  383.
7  \\
N Oph 2009 & V2672 Oph   &5060.0& 74 &  3.6 & 1021.7 & 19.6  & 17 &  3.6 &   31.
6  \\
N Pup 2004 & V574 Pup    &3330.2&100 &  9.1 & 2723.3 & 17.9  &107 &  1.7 & 2607.
7  \\ 
N Pup 2007 & V597 Pup    &4418.7&  7 & 31.1 & 1592.9 & 19.6  & 49 & 28.0 &  393.
0  \\ 
N Pup 2007b & V598 Pup   &4381.5&  5 & 68.3 & 1639.1 & 15.5  & 26 & 62.2 &  422.
2  \\ 
N Pyx 1890  & T Pyx      &5664.5& 98 &  1.5 &  399.5 & 14.3  &155 &-26.4 &  367.
5  \\ 
N Sco 1863  & U Sco      &5224.5&  0 & ---  & ---    & ---   & 44 &  1.4 &   84.
3  \\ 
N Sco 2004b & V1187 Sco  &3221.1& 56 &  5.5 & 2777.7 & 21.5  & 51 &  9.5 &  659.
5  \\ 
N Sco 2005  & V1188 Sco  &3576.8&  0 & ---  & ---    & ---   & 10 &  2.9 &   56.
7  \\ 
N Sco 2007a & V1280 Sco  &4136.4& 94 & 30.5 & 1968.3 & 10.4  &194 &  1.5 & 1939.
4  \\ 
N Sco 2008  & V1309 Sco  &4712.0& 17 & 48.5 & 1343.9 & 19.1  & 24 &  4.6 &   63.
5  \\ 
N Sco 2010 No. 2&V1311 Sco&5312.3&24 & 27.4 &  741.4 & 20.7  &  3 &  9.5 &   73.
5  \\ 
N Sco 2011  & V1312 Sco  &5713.6& 46 &  2.2 &  368.1 & 17.8  & 28 &  2.2 &  362.
2  \\ 
N Sco 2011 No. 2&V1313 Sco&5810.4&31 &  6.2 &  273.4 & 17.4  & 36 &  2.2 &  267.
4  \\
N Sct 2003  & V475 Sct   &2880.1&101 &  2.5 & 2386.8 & 19.5  & 69 &  4.4 & 1771.
6  \\ 
N Sct 2005  & V476 Sct   &3643.9&  3 & 19.6 & 2367.9 & $>$24 &  2 & 33.6 &   34.
6  \\ 
N Sct 2005b & V477 Sct   &3654.5&  1 & 24.0 & ---    & 14.6  &  2 & 23.0 &   25.
0  \\ 
N Sct 2009  & V496 Sct   &5143.9& 48 &  5.6 &  939.0 & 14.5  & 29 &  1.6 &  688.
7  \\ 
N Sgr 2002c & V4743 Sgr  &2537.9&  0 & ---  & ---    & ---   & 80 &206.0 & 1077.
6  \\ 
N Sgr 2003  & V4745 Sgr  &2755.2&  0 & ---  & ---    & ---   & 70 &  4.4 &  880  \\ 
N Sgr 2004  & V5114 Sgr  &3080.3& 49 &  2.5 & 2357.4 & 20.0  & 77 &  3.5 &  797.5  \\
N Sgr 2006  & V5117 Sgr  &3783.9& 19 &  8.0 & 2217.0 & 21.6  & 22 & 21.0 &  235.7  \\
N Sgr 2007  & V5558 Sgr  &4291.5&123 & 18.2 & 1880.5 & 13.9  &235 & 55.4 & 1876.6  \\
N Sgr 2008  & V5579 Sgr  &4575.3& 11 & 68.5 & 1506.5 & 16.6  & 15 &  9.4 & 1195.4  \\
N Sgr 2009 No. 3&V5583 Sgr&5050.0&35 &  3.7 &  991.9 & 16.5  & 25 &  4.7 &   91.5  \\
N Sgr 2009 No. 4&V5584 Sgr&5130.9&24 &  2.6 &  954.9 & 18.6  &  9 &  1.6 &  355.6  \\
N Sgr 2010 No. 2&V5586 Sgr&5310.3&19 & 29.4 &  771.5 & 17.9  &  1 &  5.4 &  ---    \\
N Sgr 2011 No. 2&V5588 Sgr&5648.3&72 & 30.5 &  437.5 & 18.0  & 54 & 35.5 &  427.5  \\ 
N TrA 2008  & NR TrA     &4558.2& 71 & 85.5 & 1547.5 & 16.3  & 58 &  5.5 & 1523.4  \\ 
CSS081007:030559+054715&HV Cet&4746.5& 19 &276.4 & 1213.1 & 19.0  & 29 & 30.2 &  130.0  \\
XMMU J115113.3-623730 & &4793.5& 49 &613.0 & 1290.1 & 15.1  & 74 &604.0 & 1282.0  \\

\enddata
\tablenotetext{a}{Reference Julian date. This is usually the time of discovery
or the time of peak observed brightness.}
\end{deluxetable}

\begin{deluxetable}{lll|rrrr|rrr}
\tablewidth{0pt}
\tablecaption{Observational Details - Recent Novae\label{tbl-obs_new}}
\tablehead{\multicolumn{2}{c}{Nova\tablenotemark{a}} & \multicolumn{1}{c}{T$_0$\tablenotemark{b}} &\multicolumn{4}{c}{Photometry} & \multicolumn{3}{c}{Spectroscopy}  \\
\colhead{}  & \colhead{}  & \colhead{}  & 
\colhead{Nights}  & \colhead{Start}  & \colhead{End} & \colhead{$V$} &
\colhead{Nights}  & \colhead{Start}  & \colhead{End}  }
\startdata
N Car 2012           & V834 Car  &5984.0&35&12.6&125.4 & 15.5 & 11 &  9.4 &  94.5 \\
PNV J13410800-5815470&N Cen 2012 &6009.9&26& 2.8& 99.7 & 14.7 & 17 &  3.9 &  95.6 \\
PNV J14250600-5845360&N Cen 2012b&6022.3&14& 9.4& 83.2 & 17.6 & 18 &  4.3 &  77.3 \\
PNV J17260708-2551454&N Oph 2012 &6012.3&25&11.3& 97.5 & 18.1 & 19 & 10.6 &  91.2 \\
PNV J17395600-2447420&N Oph 2012b&6067.0&12& 3.8& 41.8 & 15.3 & 10 &  3.9 &  36.5 \\
PNV J17452791-2305213&N Sgr 2012 &6038.5&40& 8.3& 70.3 & 15.3 & 25 &  2.4 &  67.0 \\
                     &N Sco 2012 &6085.8& 0& ---&  --- & ---  &  3 & 13.9 &  17.8 \\
PNV J17522579-2126215&N Sgr 2012c&6105.8& 0& ---&  --- & ---  &  2 &  3.8 &   8.4 \\
PNV J18110375-2717276&N Sgr 2012b&6000  & 0& ---&  --- & ---  &  2 &435.4 & 437.3 \\
\enddata
\tablenotetext{a}{Nova designations in the second column have not been authorized
by the GCVS, and are unofficial.} 
\tablenotetext{b}{Reference Julian date-2450000. This is usually the time of discovery or the time of peak observed brightness.}
\end{deluxetable}

\begin{deluxetable}{lll|rrrr|rrr}
\tablewidth{0pt}
\tablecaption{Observational Details - Novae in the LMC\label{tbl-obs_LMC}}
\tablehead{\multicolumn{2}{c}{Nova} & \multicolumn{1}{c}{T$_0$\tablenotemark{a}} &\multicolumn{4}{c}{Photometry} & \multicolumn{3}{c}{Spectroscopy} \\
\colhead{}  & \colhead{}  & \colhead{}  &
\colhead{Nights}  & \colhead{Start}  & \colhead{End} & \colhead{$V$} &
\colhead{Nights}  & \colhead{Start}  & \colhead{End} }
\startdata
N Dor 1937 & YY Dor &3298.7&167 &  1.1 & 2765.8 & 18.9  & 36 &  1.1 &  112.9 \\
N LMC 2005  &       &3696.9&152 &  7.1 & 2278.0 &$>$19.6  &105 &  7.2 & 670.2 \\
N LMC 2009  &       &4867.6&143 &  2.0 & 1198.9 & 20.1  & 53 &  2.0 &  187.3 \\
N LMC 2009b &       &4956.5& 47 & 11.0 & 1018.1 & 18.7  & 41 &  8.0 &  261.1 \\ 
N LMC 2012  &       &6012.9& 34 &  1.6 &   90.0 & 18.4  & 18 &  0.6 &   45.6 \\
\enddata
\tablenotetext{a}{Reference Julian date. This is usually the time of discovery
or the time of peak observed brightness.}
\end{deluxetable}


\small
\begin{deluxetable}{ll|lrrl}
\tablewidth{0pt}
\tablecaption{Nova Characteristics\label{tbl-meas}}

\tablehead{\multicolumn{2}{c}{Nova} & \colhead{Spec} & \colhead{FWHM(H$\alpha$)}& \colhead{day} & \colhead{Phot} \\
\colhead{} & \colhead{} & \colhead{Type} & \colhead{(\AA)} & \colhead{} & \colhead{Type\tablenotemark{a}} }
\startdata
N Aql 2005 & V1663 Aql    & Fe II & 39 & 49.0 & - \\
N Car 2008 & V679 Car     & Fe II & 44 &  4.1 & - \\
N Car 2012 & V834 Car     & Fe IIw& 41 & 76.5 & S \\
N Cen 2005 & V1047 Cen    & Fe II & 28 &  6.9 & - \\
N Cen 2007 & V1065 Cen    & Fe IIw& 45 & 13.9 & - \\
N Cen 2009 & V1213 Cen    & Fe II & 39 &125.8 & - \\
PNV J13410800-5815470&N Cen 2012 &Fe IIw&31&28.9&S,P,D\\
PNV J14250600-5845360&N Cen 2012b&Fe II&36&12.5& S\\
N Cir 2003 & DE Cir       & He-N  &133 & 12.0 & - \\
N Cru 2003 & DZ Cru       & Fe IIw& 28 & 20.6 & - \\
N Dor 1937 & YY Dor       & He-N  &145 &  5.1 & S \\
N Eri 2009 & KT Eri       & He-N  & 98 & 23.6 & P \\
N Lup 2011 & PR Lup       & Fe IIw& 31 & 19.5 & F \\
N Mus 2008 & QY Mus       & Fe IIw& 34 &185.9 & D?\\
N Nor 2005 & V382 Nor     & Fe II & 34 & 56.9 & - \\
N Nor 2007 & V390 Nor     & Fe IIw& 28 & 13.4 & - \\
N Oph 1898 & RS Oph       & He-N  & 22 & 25.6 & P \\
N Oph 2003 & V2573 Oph    & Fe IIw& 29 &822.4 & - \\
N Oph 2004 & V2574 Oph    & Fe IIw& 36 & 30.3 & - \\
N Oph 2006 & V2575 Oph    & Fe IIw& 29 & 94.0 & - \\
N Oph 2006b & V2576 Oph   & Fe IIw& 55 & 45.7 & S \\
N Oph 2007 & V2615 Oph    & Fe IIw& \tablenotemark{b} &  -   & - \\
N Oph 2008 & V2670 Oph    & Fe II & 28 & 26.3 & - \\
N Oph 2008b & V2671 Oph   & Fe II & 28 & 20.7 & - \\
N Oph 2009 & V2672 Oph    & He-N  &190 &  3.6 & S \\
PNV J17260708-2551454&N Oph 2012 &Fe IIw&20&36.6&D\\
PNV J17395600-2447420&N Oph 2012b&Fe II &66& 3.9&S\\
N Pup 2004 & V574 Pup     & Fe IIw& 51 &  9.6 & S \\
N Pup 2007 & V597 Pup     & He-N? & 79 & 31.1 & - \\
N Pup 2007b & V598 Pup    & Fe II & 50 & 68.3 & - \\
N Pyx 1890  & T Pyx       & Fe IIw& 73 &194.8 & S \\
N Sco 1863  & U Sco       & He-N  &148 &  3.4 & - \\
N Sco 2004b & V1187 Sco   & Fe IIw& 64 & 18.6 & S \\
N Sco 2005  & V1188 Sco   & Fe II & 35 & 43.8 & - \\
N Sco 2007a & V1280 Sco   & Fe IIw& 23 & 55.5 & - \\
N Sco 2008  & V1309 Sco   & Sy?   &  4 &  7.6 & - \\
N Sco 2010 No. 2&V1311 Sco& He-N? & 79 & 46.5 & S \\
N Sco 2011  & V1312 Sco   & Fe IIw& 39 &  2.2 & O \\
N Sco 2011 No. 2&V1313 Sco& He-Nw & 94 &  6.1 & S \\
N Sco 2012  &             & Fe II & 45 & 15.8 & D \\
N Sct 2003  & V475 Sct    & Fe IIw& 30 & 53.4 & F \\
N Sct 2005  & V476 Sct    & Fe II & 26 & 33.6 & - \\
N Sct 2005b & V477 Sct    & Fe II & 56 & 23.0 & - \\
N Sct 2009  & V496 Sct    & Fe IIw& 26 &215.0 & - \\
N Sgr 2002c & V4743 Sgr   & Fe II & 43 &221.9 & - \\
N Sgr 2003  & V4745 Sgr   & Fe IIw& 38 &102.5 & - \\
N Sgr 2004  & V5114 Sgr   & Fe IIw& 42 & 28.4 & S \\
N Sgr 2006  & V5117 Sgr   & Fe IIw& 35 & 49.9 & S \\
N Sgr 2007  & V5558 Sgr   & Fe IIw& 20 &141.2 & F \\
N Sgr 2008  & V5579 Sgr   & Fe IIw& 45 &1123.5& - \\
N Sgr 2009 No. 3&V5583 Sgr& Fe II & 66 &  4.7 & S \\
N Sgr 2009 No. 4&V5584 Sgr& Fe IIw& 26 & 11.6 & - \\
N Sgr 2010 No. 2&V5586 Sgr& He-N? & 84 &  5.4 & - \\
N Sgr 2011 No. 2&V5588 Sgr& Fe II & 16 & 44.6 & J \\
PNV J17452791-2305213&N Sgr 2012&He-N&132&10.4& S \\
PNV J18110375-2717276&N Sgr 2012b&\ion{Fe}{2} &8&101.6&-\\
PNV J17522579-2126215&N Sgr 2012c&He-N    & 92& 4.5& S\\
N TrA 2008  & NR TrA      &\ion{Fe}{2}w   & 19 & 29.5 & - \\
CSS081007:030559+054715&HV Cet&He-N&\tablenotemark{c}& - & - \\
XMMU J115113.3-623730&--- &\tablenotemark{d}& 24 &604.0& - \\
N LMC 2005  &             &\ion{Fe}{2}w& 20 & 20.1& F,D,C\\
N LMC 2009  &             & He-Nw      & 96 &  4.0 & P \\
N LMC 2009b &             &\ion{Fe}{2}w& 22 & 82.4 & D \\
N LMC 2012  &             & He-Nw      &171 &  2.6 & S \\
\enddata
\tablenotetext{a}{Types from visual comparison to \cite{SSH10} Figure~2.}
\tablenotetext{b}{No H$\alpha$ spectrum available.}
\tablenotetext{c}{H$\alpha$ line profile is peculiar.}
\tablenotetext{d}{First spectrum obtained too late to classify nova type.}
\end{deluxetable}


\clearpage

\normalsize

\begin{thebibliography}{}

\bibitem[Bode \& Evans(2008)]{BE08} Bode, M.F. \& Evans, A., {\it Classical Novae, 2$^{nd}$ edition}, 2008, (Cambridge)

\bibitem[Bond et al.(2003)]{iauc8185} Bond, H.E. et al.\ 2003, IAUC, 8185

\bibitem[Dworetsky(1983)]{D83} Dworetsky, M.M. 1982, \mnras, 203, 917

\bibitem[Ederoclite et al.(2006)]{E06} Ederoclite, A. et al.\ 2006 \aap, 459, 875

\bibitem[Evans et al.(2012)]{Eva12} Evans, A. et al. 2012, \mnras, in press

\bibitem[Ford(1978)]{For78} Ford, H.C. 1978, \apj, 219, 595 

\bibitem[Hachisu \& Kato(2003)]{HK03} Hachisu, I. \& Kato, M. 2003, \apj, 598, 527

\bibitem[Hamuy et al.(1992)]{Ham92} Hamuy, M., Walker, A.R., Suntzeff, N.B., Gigoux, P., Heathcote, S.R. \& Phillips, M.M. 1992, \pasp, 104, 533

\bibitem[Hamuy et al.(1994)]{Ham94} Hamuy, M., Suntzeff, N.B., Heathcote, S.R., Walker, A.R., Gigoux, P. \& Phillips, M.M. 1992, \pasp, 106, 566

\bibitem[Helton et al.(2010)]{Hel10} Helton, L.A. et al.\ 2010, \aj, 140, 1347

\bibitem[Hounsell et al.(2010)]{Hou10} Hounsell, R. et al.\ 2010, \apj, 724, 480

\bibitem[Hughes et al.(2010)]{Hug10} Hughes, J.P, et al. 2010, ATel, 2771

\bibitem[Hung, Chen \& Walter(2011)]{Hun11} Hung, L.W., Chen, W.P. \& Walter,
   F.M. 2012, ASPC, 451, 271

\bibitem[Jurdana-\u{S}epi\'c(2012)]{Jur12} Jurdana-\u{S}epi\'c, R.,
Ribeiro, V.A.R.M., Darnley, M.J., Munari, U. \& Bode, M.F. 2012, \aap, 537, A34

\bibitem[Kallman \& McCray(1980)]{KM80} Kallman, T. \& McCray, R. 1980, \apj, 242, 615

\bibitem[Kraft(1963)]{K63} Kraft, R.F. 1963, in ``Adv. Astr. Ap.'', ed. V. Kopal, 2, 43.

\bibitem[Landolt(1992)]{Landolt92} Landolt, A.U. 1992, \aj, 104, 340

\bibitem[Liller(2003)]{L03} Liller, W. 2003, IAUC, 219
\bibitem[Liller(2004)]{lil04} Liller, W. 2004, IAUC, 8422

\bibitem[Mason et al.(2010)]{Mas10} Mason, E., Diaz, M., Williams, R.E., Preston, G. \& Bensby, T. 2010, A\&A, 516, 108

\bibitem[Maxwell et al.(2012)]{Max12} Maxwell, M.P. et al. 2012, \mnras, 416, 1465

\bibitem[McClintock et al.(1975)]{MCT75} McClintock, J.E., Canizares, C.R. \& Tarter, C.B. 1975, \apj, 198, 641

\bibitem[McLaughlin(1960)]{McL60} McLaughlin, D.B. 1960, in ``Stellar Atmospheres'', ed. J.L. Greenstein, 585

\bibitem[Munari et al.(2011)]{Mun11} Munari U., Ribeiro V.A.R.M., Bode M.F. \&
Saguner T. 2011, MNRAS, 410, 525

\bibitem[Naik et al.(2010)]{N10} Naik, S., Bannerjee, D.P.K., Ashok, N.M., \& Das, R.K. 2010, \mnras, 404, 367

\bibitem[Naito et al.(2012)]{N12} Naito, H. et al.\ 2012, \aap, 543, 86

\bibitem[Nakano et al.(2012)]{Nak12} Nakano, S. et al.\ 2012, CBET, 3140

\bibitem[Oke(1990)]{Oke90} Oke, J.B. 1990, \aj, 99, 1621 

\bibitem[Patterson et al.(2010)]{Pat10} Patterson, J. et al. 2010, ATel, 2777

\bibitem[Payne-Gaposchkin(1957)]{PG57} Payne-Gaposchkin, C. 1957, ``The Galactic Novae'' (North-Holland, Amsterdam)

\bibitem[Pietsch et al.(2007)]{Pie07} Pietsch, W. et al. 2007 \aap, 465, 375


\bibitem[Read et al.(2007)]{RSE07} Read, A.M., Saxton, R.D. \& Esquej, P. 2007,
ATel 1282

\bibitem[Rushton et al.(2008)]{Rus08} Rushton, M.T,, Evans A., Eyres S.P.S., Van Loon J.T. \& Smalley B. 2008, MNRAS, 386, 289

\bibitem[Schwarz et al.(2011)]{Swift11} Schwarz, G.J. et al. 2011, \apjs, 197, 31

\bibitem[Seach et al.(2011)]{Sea11} Seach, J. et al. 2011, IAUC, 9233

\bibitem[Shafter et al.(2011)]{Sha11} Shafter, A.W., Darnley, M.J., Hornoch, K., Filippenko, A.V., Bode, M.F., Ciardullo, R., Misselt, K.A., Hounsell, R.A., Chornock, R. \& Matheson, T. 2011, \apj, 734, 12


\bibitem[Shore et al.(2011)]{SNS11} Shore, S.N., Augusteijn, T., Ederoclite, A. \& Uthas, H., \aap, 533, 8

\bibitem[Silviero et al.(2005)]{S05} Silviero, A., Munari, U. \& Jones, A.F. 2005, IBVS, 5638

\bibitem[Smak et al.(2001)]{SBZ01} Smak, J.I., Belczy\'nski, K. \& Zola, S. 2001, Acta Astron., 51, 117

\bibitem[Starrfield(1971)]{S71} Starrfield, S. 1971, \mnras, 152, 307

\bibitem[Steiner \& Diaz(1998)]{SD98} Steiner, J.E. \& Diaz, M.P. 1998, \pasp, 110, 276


\bibitem[Stringfellow \& Walter(2006)]{Str06} Stringfellow, G.S. \& Walter, F.M. 2006, \apss, 304, 401

\bibitem[Strope et al.(2010)]{SSH10} Strope, R.J, Schaefer, B.E., \&
Henden, A.A. 2010, \aj, 140, 34

\bibitem[Subasavage et al.(2010)]{S10} Subasavage, J.P et al. 2010, SPIE, 7737,  77371C

\bibitem[Walter \& Battisti(2011)]{WB11} Walter, F.M. \& Battisti, A. 2011, \baas, 217, 338.11
	
\bibitem[Williams(1991)]{Wil91} Williams, R.E., Hamuy, M., Phillips, M.M., Heathcote, S.R., Wells, L. \& Navarette, M. 1991, \apj, 376, 721

\bibitem[Williams(1992)]{Wil92} Williams, R.E. 1992, \aj, 104, 725

\bibitem[Williams et al.(1994)]{Wil94} Williams, R.E., Philips, M.M., \& Hamuy, M.
1994, \apjs, 90, 297



\end{thebibliography}
\end{document}